\documentclass[onecollarge]{svjour2}       
\usepackage{amssymb}
\usepackage{graphicx}

\def\defeq{\;\buildrel\hbox{\small def}\over{\,=\;}}

\begin{document}
\title{The flattenings of the layers of rotating planets and satellites deformed by a tidal potential}
\author{Hugo Folonier \textperiodcentered \ Sylvio Ferraz-Mello \textperiodcentered \ Konstantin V. Kholshevnikov}

\institute{H. A. Folonier and S. Ferraz-Mello \at \emph{Instituto de Astronomia Geof\'isica e Ci\^encias
Atmosf\'ericas, Universidade de S\~ao Paulo, Brasil} \\
 \email{folonier@usp.br} and \email{sylvio@iag.usp.br}\\
 \and K.V. Kholshevnikov \at \emph{St. Petersburg University, Main (Pulkovo) Astronomical Observatory RAS, St. Petersburg, Russia}\\
 \email{kvk@astro.spbu.ru}\\}

\titlerunning{The flattenings of the layers}
\authorrunning{H. Folonier \textperiodcentered \ S. Ferraz-Mello \textperiodcentered \ K.V. Kholshevnikov} 

\maketitle

\begin{abstract}

We consider the Clairaut theory of the equilibrium ellipsoidal figures for differentiated non-homogeneous  bodies in non-synchronous rotation (Tisserand, M\'ecanique 
C\'eleste, t.II, Chap. 13 and 14) adding to it a tidal deformation due to the presence of an external gravitational force. We assume that the body is 
a fluid formed by $n$ homogeneous layers of ellipsoidal shape and we calculate the external polar flattenings $\epsilon_k$, $\mu_k$ and the mean 
radius $R_k$ of each layer, or, equivalently, their semiaxes $a_k, b_k$ and $c_k$. To first order in the flattenings, the general solution can be 
written as $\epsilon_k=\mathcal{H}_k\epsilon_{h}$ and $\mu_k=\mathcal{H}_k\mu_{h}$, where $\mathcal{H}_k$ is a characteristic coefficient for each 
layer which only depends on the internal structure of the body and $\epsilon_{h}$, $\mu_{h}$ are the flattenings of the equivalent homogeneous problem. 
For the continuous case, we study the Clairaut differential equation for the flattening profile, using the Radau transformation to find the boundary 
conditions when the tidal potential is added. Finally, the theory is applied to several examples: i) a body composed of two homogeneous layers; ii) 
bodies with simple polynomial density distribution laws and iii) bodies following a polytropic pressure-density law.

\end{abstract}

\section{Introduction}

Several theories of tidal evolution, since the theory developed by Darwin in the XIX century (Darwin, 1880), are based on the figure of equilibrium of 
an inviscid tidally deformed body (see e.g. Ferraz-Mello et al., 2008; Ferraz-Mello, 2013). The addition of the viscosity to the model is done at a 
later stage, but the way it is introduced is not unique and can vary when different tidal theories are considered. 
Frequently, the adopted figure is a Jeans prolate spheroid or, if the rotation is important, a Roche triaxial ellipsoid (Chandrasekhar 1969). It is 
worth recalling that ellipsoidal figures, are excellent first approximations, but not exact figures of equilibrium (Poincar\'e, 1902; Lyapunov, 1925, 
1927). Besides, Maclaurin, Jacobi, Roche and Jeans ellipsoids are valid only for homogeneous bodies. Real celestial objects, however, are quite far 
from being homogeneous. This causes significant deviations which need to be taken into account in the astronomical applications.

The non-homogeneous problem, when we only consider the deformation by rotation, has been extensively studied. The problem of one body formed by $n$ 
rotating homogeneous spheroidal layers as well as its extension to the continuous case was studied by Clairaut (1743) (revisited by Tisserand 
(1891) and Wavre (1932)). Their works were based on the hypotheses of small deformations (linear theory for the polar flattenings) and constant 
angular velocity inside the body. The general case of homogeneous layers rotating with different angular velocities (non-linear theory) was studied by 
Montalvo et al. (1983) and Esteban and Vazquez (2001) (see Borisov et al. 2009 for a more detailed review), and was generalized to the continuous 
inviscid case by Bizyaev et al. (2014).

The case of uniformly rotating layers was studied by several authors. Kong et al. (2010) discussed the particular case of a body formed by two 
homogeneous layers with same angular velocity. Hubbard (2013), with a recursive numerical form of the potential of a N-layers rotating planet, in 
hydrostatic equilibrium, showed a solution for the spheroidal shapes of the interfaces of the layers.

When the tidal forces acting on the body are taken into account along with the rotation, the literature is much less extensive. Usually the spin-orbit 
synchronism is assumed, so that the rotating body solution can be used (e.g. Van Hoolst et al. (2008)). Tricarico (2014), assuming synchronism, found 
a recursive analytical solution for the shape of a body formed by an arbitrary number of layers. For this, he developed the potentials of homogeneous 
ellipsoids in terms of the polar and equatorial shape eccentricities. However, the results do not include tidally deformed bodies whose 
rotation is non-synchronous, as, for instance, the Earth, solar type stars hosting close-in planets and hot Jupiters in highly eccentric orbits.

In this work, we generalize the linear Clairaut theory, adding a tidal potential due to the presence of an external body, \emph{without the 
synchronism hypothesis}. The layout of the paper is as follows: In Section 2 we present the $2n$ classical equations of equilibrium. The resolution of 
the system of equilibrium equations is shown in Section 3. In Section 4, we study the Clairaut's equation for the continuous problem an its solution. In Section 5 we 
calculate the potential at a point in the space due to the deformed body and we calculate a generalized Love number for the differentiated non-homogeneous 
 bodies. In Section 6, we apply the theory to a body composed of two homogeneous layers, while bodies with continuous density laws are 
studied in Section 7. Finally, in the Section 8 we present the conclusions.

\section{Equilibrium equation of a fluid in rotation}

\begin{figure}[here]
\begin{center}
\includegraphics[scale=0.4]{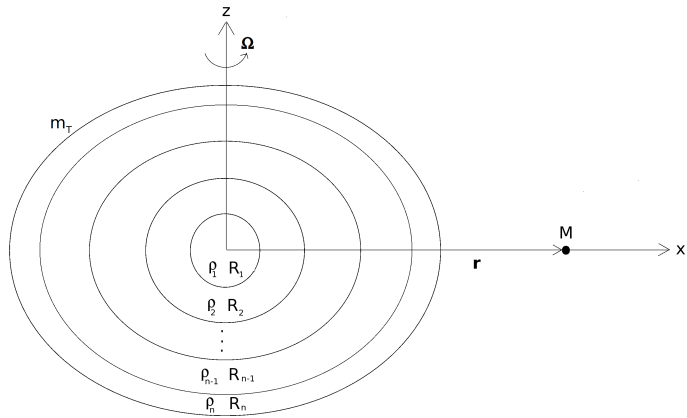}
\caption{Body of mass $m_T$ formed of $n$ homogeneous layers of density $\rho_k$ and mean radius $R_k$ rotating with angular velocity 
$\vec{\Omega}=\Omega\hat{z}$ and a point mass $M$ orbiting at a distance $r$ from its center in a plane perpendicular to the rotation axis.}
\label{fig1}
\end{center}
\end{figure}

We consider a rotating inviscid fluid of mass $m_T$ and a mass point $M$ orbiting at a distance $r$ from the center of the primary in a plane 
perpendicular to the rotation axis. We assume that the fluid is composed of $n$ homogeneous layers of density $\rho_k$ ($k=1,\cdots,n$), that 
each layer has an ellipsoidal shape with external semiaxes $a_k$, $b_k$ and $c_k$ along the coordinate axes, and angular velocity $\vec{\Omega}$. 
We define the mean radius of each layer as $R_k=\sqrt[3]{a_kb_kc_k}$. We choose a reference system such that $\vec{r}=r\hat{x}$ and 
$\vec{\Omega}=\Omega\hat{z}$ where $\hat{x}$, $\hat{z}$ are unit vectors along the axes $x$, $z$ (Figure \ref{fig1}).

Now, if we consider one point on the surface of the $\ell$-th layer, with position vector 
$\vec{x}_\ell=\xi_\ell\hat{x}+\eta_\ell\hat{y}+\zeta_\ell\hat{z}$ and velocity is $\vec{v}_\ell=\vec{\Omega}\times\vec{x}_\ell$, 
we can use the same equation used in the study of equilibrium ellipsoids (see Tisserand, 1891, Chap. 8 and 13; Jeans, 1929, Sec. 215-216; 
Jardetzky, 1958; Chandrasekhar, 1969), which expresses the fact that the total force acting on a point of its surface must be perpendicular to 
the surface
\begin{equation}
\nabla_\ell \Phi_\ell\propto \nabla_\ell V_G + \vec{\Omega}\times(\vec{\Omega}\times\vec{x}_\ell)
\end{equation}
where
\begin{equation}
 \Phi_\ell(\xi_\ell,\eta_\ell,\zeta_\ell) = \frac{\xi_\ell^2}{a_\ell^2}+\frac{\eta_\ell^2}{b_\ell^2}+\frac{\zeta_\ell^2}{c_\ell^2}-1=0
\end{equation}
is the equation of the surface of the ellipsoid, $V_G$ is the potential of the gravitational forces at $\vec{x}_\ell$ and the last term 
corresponds to the centripetal acceleration. The use of the above equilibrium equation in a case where the tidal force field is changing because of 
the external body needs a justification. Eqn. (1) means that no change in the shape of the body occurs because of internal forces; the shape will 
change, but only because of the relative change of the position of the external body.

Hence, we obtain the equilibrium equations 
\begin{eqnarray}
 \Omega^2 &=&  \displaystyle\frac{1}{\xi_\ell}\displaystyle\frac{\partial V_G}{\partial \xi_\ell}   - \displaystyle\frac{\alpha_\ell}{\zeta_\ell}\displaystyle\frac{\partial V_G}{\partial \zeta_\ell}\nonumber\\
 \Omega^2 &=&  \displaystyle\frac{1}{\eta_\ell}\displaystyle\frac{\partial V_G}{\partial \eta_\ell} - \displaystyle\frac{\beta_\ell}{\zeta_\ell}\displaystyle\frac{\partial V_G}{\partial \zeta_\ell},
\label{eq:equilibrio1}
\end{eqnarray}
where
\begin{equation}
 \alpha_\ell=\displaystyle\frac{c_\ell^2}{a_\ell^2}<1,\ \ \ \ \beta_\ell=\displaystyle\frac{c_\ell^2}{b_\ell^2}<1.
\end{equation}

The problem of finding the equilibrium figure (i.e. the values of the semiaxes $a_k$, $b_k$ and $c_k$) is equivalent to finding the $2n$ external polar 
flattenings
\begin{eqnarray}
 \epsilon_k &=&\frac{a_k-c_k}{a_k} \approx \frac{1-\alpha_k}{2}\nonumber\\
 \mu_k      &=&\frac{b_k-c_k}{b_k} \approx \frac{1-\beta_k}{2}
\end{eqnarray}
for each layer. For this, we will use the $2n$ equilibrium equations (\ref{eq:equilibrio1}).

The gravitational potential can be written as the sum of the potential due to the mass $M$ and the sum of the potentials of each layer. It can also be 
written as the sum of the potentials of $n$ superposed homogeneous ellipsoids, with semiaxes $a_k$, $b_k$ and $c_k$ and densities
\begin{equation}
 \sigma_k = \rho_k-\rho_{k+1},
\label{eq:density}
\end{equation}
with $\rho_{n+1}=0$. The mass of each partial ellipsoid is
\begin{equation}
 m_k = \frac{4\pi}{3}\sigma_k R^3_k.
\end{equation}

If we call $V_k$ the potential of each ellipsoid, the total potential is
\begin{equation}
 V_G = V_{tid} + \sum_{k=1}^{n} V_k.
\label{eq:pot.eli}
\end{equation}

As the equilibrium equations (\ref{eq:equilibrio1}) are linear in $V$, we can write
\begin{equation}
 \Omega^2=\chi_\ell^{(i)}(V_{tid}) + \sum_{k=1}^{n} \chi_\ell^{(i)}(V_k),
\label{eq:equi}
\end{equation}
where $\chi_\ell^{(1)}$ and $\chi_\ell^{(2)}$ are the operators
\begin{eqnarray}
 \chi_\ell^{(1)} &=&  \displaystyle\frac{1}{\xi_\ell} \frac{\partial }{\partial \xi_\ell}  - \displaystyle\frac{\alpha_\ell}{\zeta_\ell}\displaystyle\frac{\partial }{\partial \zeta_\ell}\nonumber\\
 \chi_\ell^{(2)} &=&  \displaystyle\frac{1}{\eta_\ell}\frac{\partial }{\partial \eta_\ell} - \displaystyle\frac{\beta_\ell}{\zeta_\ell}\displaystyle\frac{\partial }{\partial \zeta_\ell}.
\label{eq:equi.vk}
\end{eqnarray}

\section{Flattenings of the layers}

The next step is to calculate the contribution of each potential to the equilibrium equations (\ref{eq:equi}). If we consider the 
contributions to the potentials due to the inner and outer layers on the $\ell$-th layer, we obtain the equations
\begin{eqnarray}
  \Omega^2 &=& -\frac{3GM}{r^3} + \sum_{k=1}^{\ell-1} \frac{Gm_k}{R_\ell^3} \left[2\epsilon_\ell - \frac{6\epsilon_k}{5}\left(\frac{R_k}{R_\ell}\right)^2\right] + \frac{Gm_\ell}{R_\ell^3} \frac{4\epsilon_\ell}{5} + \sum_{k=\ell+1}^{n} \frac{Gm_k}{R_k^3} \left[2\epsilon_\ell-\frac{6\epsilon_k}{5}\right]\nonumber\\
 \Omega^2 &=& \sum_{k=1}^{\ell-1} \frac{Gm_k}{R_\ell^3} \left[2\mu_\ell - \frac{6\mu_k}{5}\left(\frac{R_k}{R_\ell}\right)^2\right] + \frac{Gm_\ell}{R_\ell^3} \frac{4\mu_\ell}{5} + \sum_{k=\ell+1}^{n} \frac{Gm_k}{R_k^3} \left[2\mu_\ell-\frac{6\mu_k}{5}\right],
\end{eqnarray}
(see Appendix A in the Online Supplement), which can be solved with respect to the $\ell$-th layer flattenings, giving
\begin{eqnarray}
 \gamma_\ell \epsilon_\ell &=& (\epsilon_J+\epsilon_M) \left(\frac{R_\ell}{R_n}\right)^3 + \displaystyle\sum_{k=1}^{\ell-1} \alpha_{\ell k}\epsilon_k + \displaystyle\sum_{k=\ell+1}^{n} \beta_{\ell k}\epsilon_k\nonumber\\
 \gamma_\ell \mu_\ell &=& \epsilon_M \left(\frac{R_\ell}{R_n}\right)^3 + \displaystyle\sum_{k=1}^{\ell-1} \alpha_{\ell k}\mu_k      + \displaystyle\sum_{k=\ell+1}^{n} \beta_{\ell k}\mu_k,
\label{eq:achata1}
\end{eqnarray}
where $\epsilon_M$ and $\epsilon_J$ are the flattenings of the equivalent MacLaurin and Jeans homogeneous spheroids:
\begin{equation}
\epsilon_M = \frac{5R_n^3\Omega^2}{4m_TG} \ \ \ \ \ \ \ \ \ \ \epsilon_J = \frac{15MR_n^3}{4m_Tr^3},
\label{def:em}
\end{equation}
where $m_T$ is the mass of the body.

These flattenings are obtained as solutions of the rotational and tidal problem, respectively, when the deformed body is a homogeneous 
spheroid with the same mass $m_T$ and mean radius $R_n$ as the considered body (see online Appendix B). The coefficients 
$\alpha_{\ell k}$, $\beta_{\ell k}$ and $\gamma_{\ell}$ are
\begin{eqnarray}
 \alpha_{\ell k} &=& \frac{3m_k}{2m_T}\left(\frac{R_k}{R_\ell}\right)^2 \nonumber\\
 \beta_{\ell k}  &=& \frac{3m_k}{2m_T}\left(\frac{R_\ell}{R_k}\right)^3\nonumber\\
 \gamma_\ell     &=& 1 + \frac{3(m_T-m_\ell)}{2m_T} - \sum_{k=\ell+1}^{n} \frac{5m_k}{2m_T} \frac{(R_k^3-R_\ell^3)}{R_k^3}.
\label{eq.para}
\end{eqnarray}

If we divide the equations for $\epsilon_\ell$ by $\epsilon_J+\epsilon_M$ and divide the equations for $\mu_\ell$ by $\epsilon_M$, we obtain the same 
equation for the two polar flattenings:
\begin{equation}
 \gamma_\ell \mathcal{H}_\ell =  \left(\frac{R_\ell}{R_n}\right)^3 + \displaystyle\sum_{k=1}^{\ell-1} \alpha_{\ell k}\mathcal{H}_k + \displaystyle\sum_{k=\ell+1}^{n} \beta_{\ell k}\mathcal{H}_k,
\label{eq:achata2}
\end{equation}
where 
\begin{equation}
\mathcal{H}_j  \defeq \frac{\epsilon_j}{\epsilon_J+\epsilon_M} = \frac{\mu_j}{\epsilon_M} \ \ \ \ \ \ \ \ (j=1,\cdots,n).
\end{equation}

It is worth emphasizing that these equations naturally associate the flattening of the homogeneous MacLaurin spheroid with the polar 
flattenings $\mu_k$ calculated using the minor semi-axis of the tidally deformed equator. This is so because the tide also acts shortening the polar 
axis. While the flattenings $\epsilon_k$ increase because of the tide, the flattenings $\mu_k$ remain the same as in absence of tide. Therefore, the 
tide increases the  mean polar flattening of the layers. The 3 axes of the layer are  $a_j=R_j[1+\mathcal{H}_j(\epsilon_M+2\epsilon_J)/3]$; 
$b_j=R_j[1+\mathcal{H}_j(\epsilon_M-\epsilon_J)/3]$; $c_j=R_j[1+\mathcal{H}_j(-2\epsilon_M-\epsilon_J)/3].$

It is important to note that if the orbital motion is synchronous with the rotation, in which case $\epsilon_J=3\epsilon_M$, the system 
(\ref{eq:achata1}) is completely equivalent to that found by Tricarico (2014), where the square of the polar and equatorial ``\emph{eccentricities}'' used there 
are related to the polar flattenings through $e_{pi}^2\approx2\epsilon_i$ and $e_{qi}^2\approx2\epsilon_i-2\mu_i$.

The calculations done are valid only for small flattenings, i.e. they assume that the perturbation due to the tide and the rotation are small enough 
so as not to deform too much the body (in the second order, the figure ceases to be an ellipsoid).

\section{Extension to the continuous case} \label{continuous}

In order to extend to the continuous case (following Tisserand, 1891, Chap. 14\footnote{see Appendix \ref{sec:appC} in the Online Supplement 
for more details.}), we assume that the number of layers tends to infinity so that the increments $\Delta R_k=R_k-R_{k-1}$ are infinitesimal quantities. 
When $\Delta R_k\rightarrow 0$, the equation (\ref{eq:achata2}) becomes
\begin{equation}
 \frac{5x^2}{3}f(x) \mathcal{H}(x) = \frac{2f_n}{3} x^5 + \int_{z=0}^{z=x} \widehat{\rho}(z) d(z^5\mathcal{H}(z)) + x^5 \int_{z=x}^{z=1} \widehat{\rho}(z) d\mathcal{H}(z),
\label{eq:dif1}
\end{equation}
where $x=R/R_n$ is the normalized mean radius ($x(0)=0$ in the center and $x(R_n)=1$ 
on the surface), $\widehat{\rho}(x)=\rho(R)/\rho_0$ is the normalized density ($\rho_0$ is the density in the center, therefore $\widehat{\rho}(0)=1$)
and the function $f(x)$ is
\begin{equation}
 f(x)=\displaystyle3\int_0^{x}\widehat{\rho}(z)z^2dz,
\end{equation}
with $f(0)=0$ and $f(1)=f_n$.

Deriving (\ref{eq:dif1}) with respect to $x$, we have
\begin{equation}
 \frac{2f(x)}{3x^3}\mathcal{H}(x)+\frac{f(x)}{3x^2}\mathcal{H}'(x) = \frac{2f_n}{3} +  \int_{z=x}^{z=1} \widehat{\rho}(z)d\mathcal{H}(z),
\label{eq:dif2}
\end{equation}
and deriving once more we obtain the differential equation for the flattening profile
\begin{equation}
 \mathcal{H}''(x)+ \frac{6\widehat{\rho}(x) x^2}{f(x)} \mathcal{H}'(x) + \left(\frac{6\widehat{\rho}(x) x}{f(x)}-\frac{6}{x^2}\right) \mathcal{H}(x) = 0.
\label{eq:dif3}
\end{equation}

It is a homogeneous linear differential equation of second order with non constant coefficients and it turns out to be the same for both flattenings. 
It is the same expression found by Clairaut (Jeffreys, 1953).

The Eq. (\ref{eq:dif1}) allows us to calculate easily the limits $\mathcal{H}_n$ that the proportionality parameter $\mathcal{H}$ can take at the 
surface. In the homogeneous case $\widehat{\rho}(x)=1$, the integrals can be calculated trivially. At the surface $x=1$, we obtain $\mathcal{H}_n=1$. 
In the non-homogeneous case, if the density is a non-increasing function ($\frac{d\widehat{\rho}}{dx}\leqslant 0$), we have, at the surface
\begin{eqnarray}
 \mathcal{H}_n &=& \frac{2}{5} + \frac{3}{5f_n}\int_{z=0}^{z=1} \widehat{\rho}(z) d(z^5\mathcal{H}(z))\nonumber\\
               &=& \frac{2}{5} + \frac{3}{5f_n}\left[\widehat{\rho}_n\mathcal{H}_n-\int_{z=0}^{z=1}z^5\mathcal{H}(z)d\widehat{\rho}(z)\right]\geqslant\frac{2}{5}.
\label{eq:h<0.4}
\end{eqnarray}
Then, under the assumption of equilibrium, a non-homogeneous body will have flattenings on the surface with values between 0.4 and 1 times the values 
they would have if the body was homogeneous.

\subsection{\textit{Boundary conditions. Radau transformation}}
\label{radau}

The differential equation (\ref{eq:dif3}) requires two boundary conditions to be solved. However, before attempting to find these boundary 
conditions, we will show two relationships that will prove useful later. The first relationship is obtained from equation (\ref{eq:dif2}), where at 
$x=1$ we have
\begin{equation}
\mathcal{H}'_n = 2(1-\mathcal{H}_n).
\label{eq:relacion1}
\end{equation}
The second relationship is obtained from the differential equation (\ref{eq:dif3}) by just multiplying it by $f(x)/x^2$ and evaluating the 
resulting equation in the neighborhood of $x=0$ (Note that $f(x)\sim x^3 + 3(d\widehat{\rho}/dx)_0x^4/4$ and $\widehat{\rho}(x)\sim 1 + (d\widehat{\rho}/dx)_0x$). It is
\begin{equation}
\mathcal{H}'_0 = -\frac{d\widehat{\rho_0}}{dx}\frac{ \mathcal{H}_0}{4},
\label{eq:relacion2}
\end{equation}
where $\frac{d\widehat{\rho_0}}{dx}$ is the derivative of the density at $x=0$.

In practical applications (see Section 6), it is convenient to introduce the Radau transformation
\begin{equation}
 \eta(x) = \frac{x\mathcal{H}'(x)}{\mathcal{H}(x)},
\label{eq:tran-radau}
\end{equation}
and rewritten Clairaut's equation as the Ricatti differential equation
\begin{equation}
 \eta' + \frac{\eta^2}{x}+\left[q(x)+\frac{5}{x}\right] \eta+ q(x) = 0,
\label{eq:radau}
\end{equation}
where
\begin{equation}
 q(x) \defeq \frac{6}{x}\left(\frac{\widehat{\rho}(x)x^3}{f(x)}-1\right).
\end{equation}

In the new variables the boundary condition is
\begin{equation}
 \eta(x=0) = 0.
\end{equation}
The variable $\eta$ is sometimes referred to as \emph{Radau's parameter} (Bullen, 1975). Defining $\eta(x=1)=\eta_n$ and using the 
relationship (\ref{eq:relacion1}) and the transformation (\ref{eq:tran-radau}), the boundary conditions of (\ref{eq:dif3}) are
\begin{equation}
\displaystyle\mathcal{H}_n = \frac{2}{2+\eta_n} \ \ \ \ \ \ \ \ \ \mathcal{H}'_n = \frac{2\eta_n}{2+\eta_n}.
\label{eq:radau-hn}
\end{equation}
As a result of this relationship, if considering that $0.4 <\mathcal{H}_n<1$, we recover the classical result $0<\eta_n<3$ (Tisserand, 1891).

Finally, it should be noted that once $\eta(x)$ is found, we may find the profile flattening from equation (\ref{eq:tran-radau}), whose solution 
is
\begin{equation}
 \mathcal{H}(x) =  \mathcal{H}_n  e^{\int_1^x \eta(z)/z\ dz}.
\end{equation}

\section{Potential of the tidally deformed body}
\label{love}

The contribution of the $k$-th ellipsoid to the potential at an external point $\vec{x}=x\hat{x}+y\hat{y}+z\hat{z}$ is given by 
\begin{equation}
\delta V_2^{(k)} =  - \frac{Gm_kR_k^2\epsilon_k}{5r^{*3}}\left(3\cos^2\Psi_1-1\right) - \frac{Gm_kR_k^2\mu_k}{5r^{*3}}\left(3\cos^2\Psi_2-1\right),
\end{equation}
where $r^{*}=|\vec{x}|$ and $\Psi_1$ and $\Psi_2$ are the angles between the direction of the point where the potential is taken and the coordinate 
axes $x$ and $y$ respectively. \footnote{For the details of the calculation of $\delta V_2^{k}$, see Eq. (\ref{eq:Vext}) in Appendix 
A.1 (in the Online Supplement).} The total potential is the sum of the potentials of all ellipsoids:
\begin{equation}
V = -\displaystyle\frac{Gm_T}{r^*} - \frac{2k_{f}Gm_TR_n^2\epsilon_h}{15r^{*3}}\left(3\cos^2\Psi_1-1\right) - \frac{2k_{f}Gm_TR_n^2\mu_h}{15r^{*3}}\left(3\cos^2\Psi_2-1\right)
\label{eq:Vdis1}
\end{equation}
where $\epsilon_h$, $\mu_h$ are the flattenings of the equivalent homogeneous ellipsoid and the constant $k_f$ is often called fluid Love number 
(Munk and MacDonald, 1960; Correia and Rodr\'iguez, 2013). For a non-homogeneous body, we find
\begin{eqnarray}
 k_{f} \defeq\ \frac{3}{2}\frac{\sum_{k=1}^nm_kR_k^2\mathcal{H}_k}{m_TR_n^2},
\label{eq:ekf-dis}
\end{eqnarray}
or using the continuous model,
\begin{eqnarray}
 k_{f} = \frac{3}{2f_n}\int_{z=0}^{z=1}\widehat{\rho}(z)d(z^5\mathcal{H}(z)).
\label{eq:ekf-cont}
\end{eqnarray}

Using the integral form of Clairaut's equation (\ref{eq:dif2}) to evaluate the integral, we have
 \begin{eqnarray}
 k_{f} =\frac{5}{2}\mathcal{H}_n-1,
\label{eq:ekf-con}
\end{eqnarray}
which shows the link of the fluid Love number with the coefficient $\mathcal{H}_n$. This relationship is based on the fact that both constants depend 
solely on the internal structure, characterizing the inhomogeneity of the body. In the homogeneous case $\mathcal{H}_n=1$ thus recovering the 
classical result $k_f=1.5$.

\section{Two-layer Core-Shell model}
\begin{figure}[t]
\begin{center}
 \includegraphics[height=5cm,clip=]{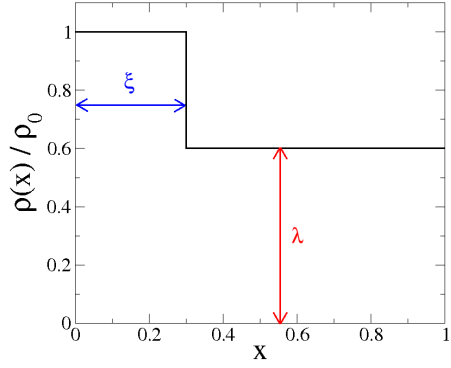}
\caption{Density profile of a body formed by two homogeneous layers. $\xi$ is the mean outer radius of the core relative to the mean outer 
radius of the shell $R_2$. $\lambda$ is the shell density relative to the core density $\rho_1$.}
\label{fig2}
\end{center}
\end{figure}

In this section we consider the simple case of a body formed of two homogeneous layers: a core with density $\rho_1$ and mean outer radius $R_1$, 
and a shell with density $\rho_2=\lambda\rho_1$ (with $\lambda<1$) and mean outer radius $R_2$ (Fig. \ref{fig2}). The densities of the superposed 
ellipsoids are
\begin{eqnarray}
 \sigma_1 &=& \rho_1(1-\lambda) \nonumber\\
 \sigma_2 &=& \rho_1\lambda,
\end{eqnarray}
and their masses are
\begin{eqnarray}
 m_1 &=& \frac{4\pi}{3}\rho_1R_2^3(1-\lambda)\xi^3 \nonumber\\
 m_2 &=& \frac{4\pi}{3}\rho_1R_2^3\lambda,
\end{eqnarray}
where $\xi=R_1/R_2$. The total mass is
\begin{equation}
 m_T = \frac{4\pi}{3}\rho_1R_2^3\Big[\lambda+(1-\lambda)\xi^3\Big].
\end{equation}
 
The polar flattenings are such that
\begin{eqnarray}
 \gamma_1\mathcal{H}_1 &=& \displaystyle\frac{R_1^3}{R_2^3}+\beta_{12}\mathcal{H}_2 \nonumber\\
 \gamma_2\mathcal{H}_2 &=& 1 +\alpha_{21}\mathcal{H}_1,
\end{eqnarray}
(see Eq. (\ref{eq:achata2})), where the coefficients are given by Eqs. (\ref{eq.para})
\begin{eqnarray}
 \alpha_{21}    &=& \frac{3m_1}{2m_T}\left(\frac{R_1}{R_2}\right)^2 = \frac{3(1-\lambda)\xi^5}{2\lambda+2(1-\lambda) \xi^3} \nonumber\\
 \beta_{12}     &=& \frac{3m_2}{2m_T}\left(\frac{R_1}{R_2}\right)^3 = \frac{3\lambda \xi^3}{2\lambda+2(1-\lambda) \xi^3} \nonumber\\
 \gamma_1       &=& 1 + \frac{3(m_T-m_1)}{2m_T} - \frac{5m_2}{2m_T} \frac{R_2^3-R_1^3}{R_2^3} = \frac{(2+3\lambda)\xi^3}{2\lambda+2(1-\lambda) \xi^3} \nonumber\\
 \gamma_2       &=& 1 + \frac{3(m_T-m_2)}{2m_T} = \frac{2\lambda+5(1-\lambda)\xi^3}{2\lambda+2(1-\lambda) \xi^3}.
\end{eqnarray}

Hence
\begin{eqnarray}
\mathcal{H}_1 &=& \frac{10\Big(\lambda+(1-\lambda)\xi^3\Big)^2}{\Big(2+3\lambda\Big)\Big(2\lambda+5(1-\lambda)\xi^3\Big)-9\lambda(1-\lambda)\xi^5}\nonumber\\
\mathcal{H}_2 &=& \frac{2\Big(\lambda+(1-\lambda)\xi^3\Big)\Big(2+3\lambda+3(1-\lambda)\xi^5\Big)}{\Big(2+3\lambda\Big)\Big(2\lambda+5(1-\lambda)\xi^3\Big)-9\lambda(1-\lambda)\xi^5}.
\label{eq:H-two-layers}
\end{eqnarray}

\begin{figure}[t]

\centerline{\hbox{
\includegraphics[height=5cm,clip=]{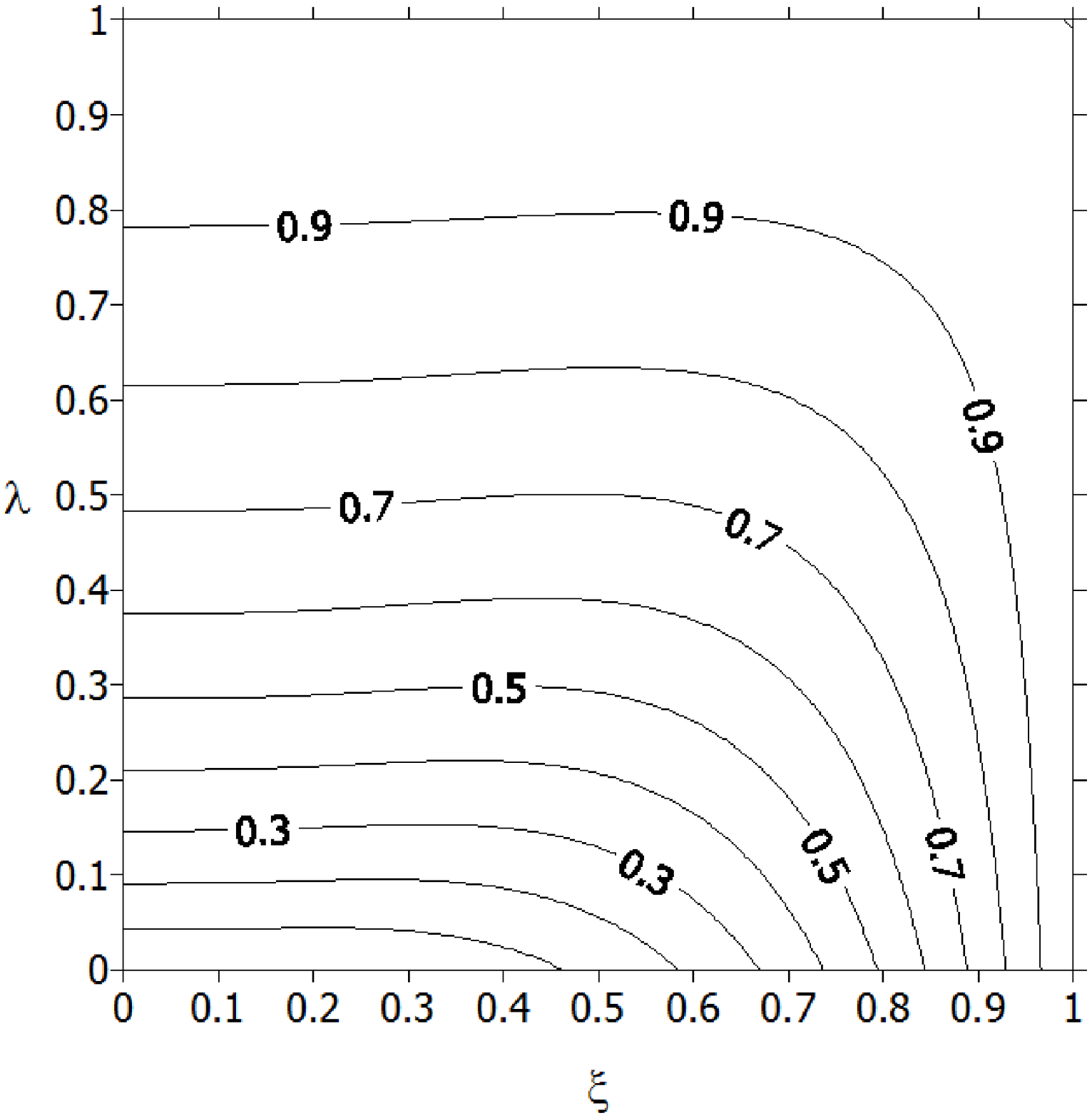}\hspace{3mm}
\includegraphics[height=5cm,clip=]{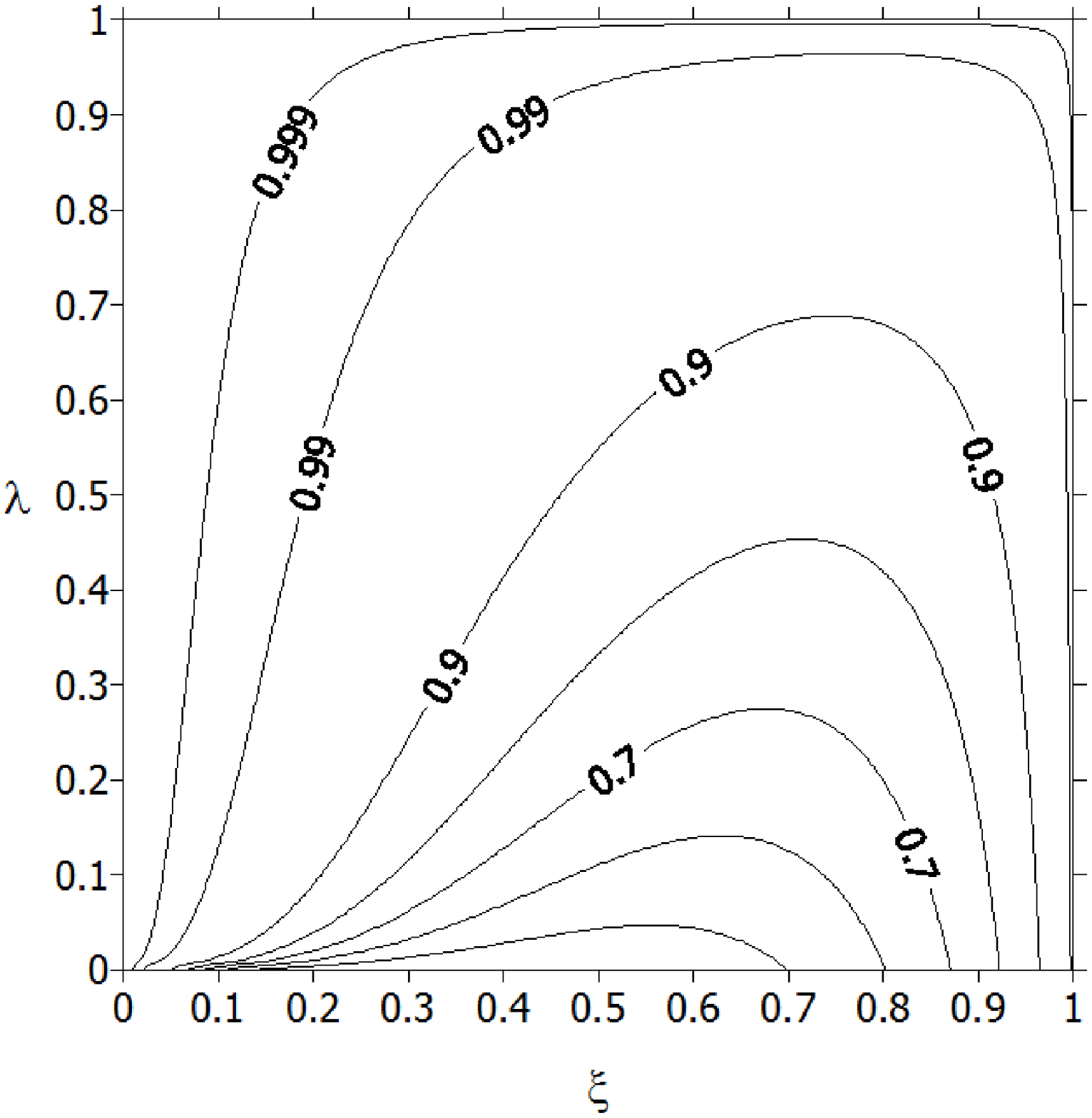}
}}
\begin{center}
\textbf{Fig. \ref{fig3}.a}\ \ \ \ \ \ \ \ \ \ \ \ \ \ \ \ \ \ \ \ \ \ \ \ \ \ \ \ \ \ \ \ \ \ \ \ \ \ \ \ \ \ \ \ \ \ \ \ \ \ \ \ \ \ \ \ \ \ \ \ \ \ \ \ \textbf{Fig. \ref{fig3}.b}\end{center}

\caption{Possible values of $\mathcal{H}_1$ (core) and $\mathcal{H}_2$ (shell) as functions of the core size $\xi$ and of the relative density of the shell 
$\lambda$.}
\label{fig3}
\end{figure}

Fig. (\ref{fig3}) shows the results obtained for the constants $\mathcal{H}_1$ and $\mathcal{H}_2$. We see that:
\begin{itemize}
 \item If $\lambda=1$ or $\xi=1$ the constants are $\mathcal{H}_1=\mathcal{H}_2$ solutions for a homogeneous body.
 \item When the core is denser than the mantle, $\mathcal{H}_2\geqslant\mathcal{H}_1$ and the flattenings of the nucleus are smaller than the 
flattenings of the surface (where $\epsilon_1=H_1(\epsilon_J+\epsilon_M) \leqslant \epsilon_2=H_2(\epsilon_J+\epsilon_M)$ and 
$\mu_1=H_1\epsilon_M\leqslant \mu_2=H_2\epsilon_M$).
 \item Since $\mathcal{H}_2\leqslant 1$, the maximum surface flattening is given by the homogeneous solution. In presence of a core, the surface 
is always less flattened than it is in the homogeneous case.
 \item While $\mathcal{H}_1$ may take all possible values between 0 and 1, $\mathcal{H}_2$ is always larger than the critical limit 0.4, 
corresponding to the degenerate limit case in which the whole mass would tend to concentrate in the center and would be surrounded by a zero-density 
shell (case of Huygens-Roche). Therefore the flattenings of the outer surface can never be less than 40\% of the homogeneous reference values. This 
is the same result given by Eq. (\ref{eq:h<0.4}) for the continuous case.
\end{itemize}

\subsection{Fluid Love number}

Using equation (\ref{eq:ekf-con}), together with the expression for $\mathcal{H}_2$ (Eqn. \ref{eq:H-two-layers}), the expression of the fluid Love number $k_f$ is
\begin{equation}
 k_f = \frac{5\Big(\lambda+(1-\lambda)\xi^3\Big)\Big(2+3\lambda+3(1-\lambda)\xi^5\Big)}{\Big(2+3\lambda\Big)\Big(2\lambda+5(1-\lambda)\xi^3\Big)-9\lambda(1-\lambda)\xi^5}-1.
\label{eq:kf-two-layers}
\end{equation}

\begin{figure}[t]
\begin{center}
\includegraphics[height=5cm,clip=]{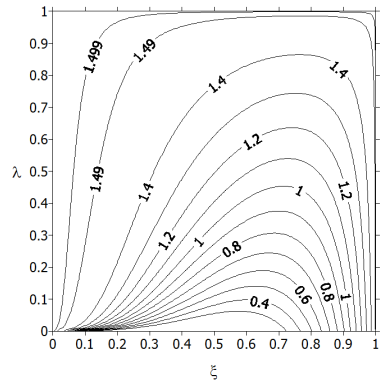}
\caption{Possible values of $k_f$ as functions of the core size $\xi$ and of relative density of the shell $\lambda$.}
\label{fig4}
\end{center}
\end{figure}

Figure (\ref{fig4}) shows the possible value of $k_f$ as a function of the core size $\xi$ and of the relative density of the shell $\lambda$. If we 
obtain $k_f$, for example by determining $\mathcal{H}_n$ by direct observation of the surface flattenings, then equation (\ref{eq:kf-two-layers}) 
defines a continuous curve of possible values for the size of the nucleus $\xi$ and the relative density of the shell $\lambda$ under the hypothesis 
of two homogeneous layers. Moreover, as can be seen in this figure, it allows us to predict a maximum value for these physical parameters.

\section{Application to different density distribution laws}

In this section, we present some applications of the theory developed in this paper to bodies with continuous density distributions. For this 
we use two examples of density distributions: polynomial and polytropic density laws.

In both cases the Clairaut's equation is solved numerically after introduction of the variable defined by the Eq. (\ref{eq:tran-radau}). The 
flattening profile $\mathcal{H}(x)$ and the Love number are then obtained through the inverse transformation.

\subsection{Polynomial density functions}

\begin{figure} [here]

\centerline{\hbox{
\includegraphics[height=5cm,clip=]{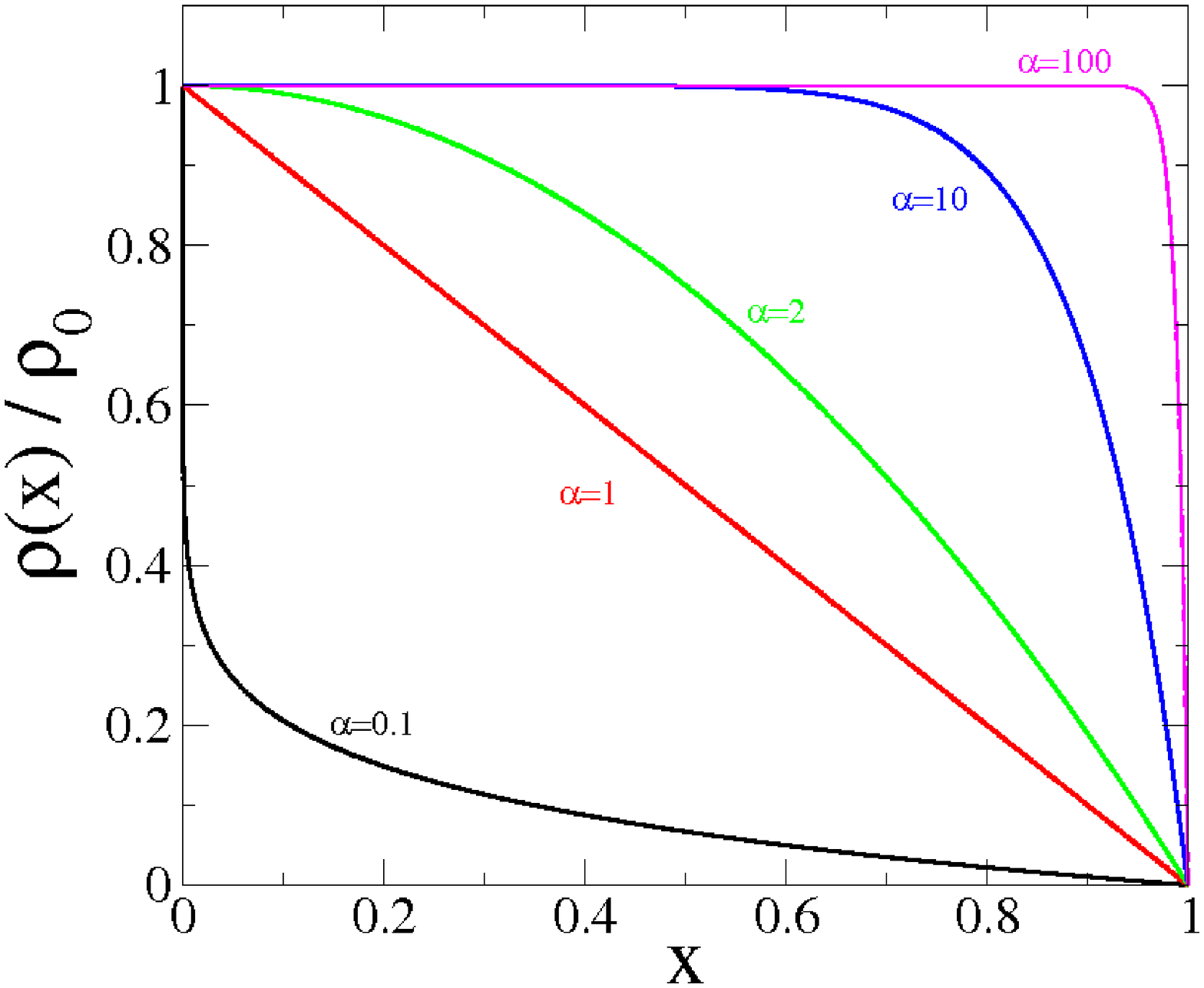}\hspace{3mm}
\includegraphics[height=5cm,clip=]{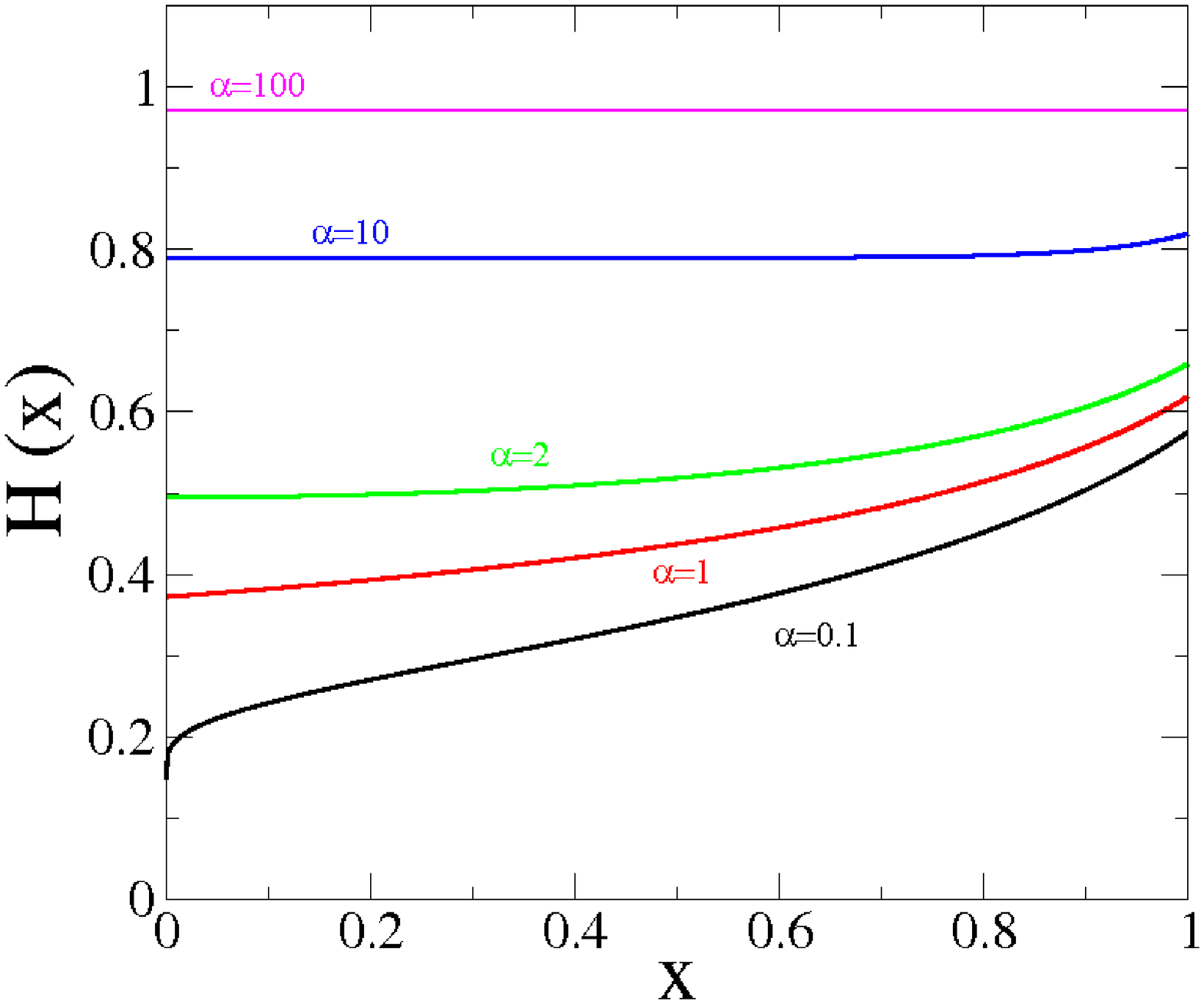}
}}
\begin{center}
\textbf{Fig. \ref{fig5}.a}\ \ \ \ \ \ \ \ \ \ \ \ \ \ \ \ \ \ \ \ \ \ \ \ \ \ \ \ \ \ \ \ \ \ \ \ \ \ \ \ \ \ \ \ \ \ \ \ \ \ \ \ \ \ \ \ \ \ \ \ \ \ \ \ \textbf{Fig. \ref{fig5}.b}\end{center}

\caption{\textbf{(a)}: Density profiles for polynomial density distributions with different values of $\alpha$. \textbf{(b)}: Flattening profile $\mathcal{H}(x)$ for the same density 
laws. $\alpha=0.1$ (black), $\alpha=1$ (red), $\alpha=2$ (green), $\alpha=10$ (blue) and $\alpha=100$ (magenta).}
\label{fig5}

\end{figure}

We consider initially a simple polynomial density law:
\begin{equation}
\widehat{\rho}(x)=1-x^\alpha,
\end{equation}
where $\alpha>0$. Figure (\ref{fig5}.a) shows the density functions for $\alpha=0.1,1,2,10$ and $100$ as functions of the normalized mean 
radius $x$.

The resulting flattening profiles $\mathcal{H}(x)$ are shown in Figure (\ref{fig5}.b). In all cases, the flattening profile $\mathcal{H}(x)$ 
is an increasing monotonic function and for all $x$, the values of $\mathcal{H}(x)$ increase when the power $\alpha$ increases.
 
\begin{figure}[h]
\begin{center}
\includegraphics[height=5cm,clip=]{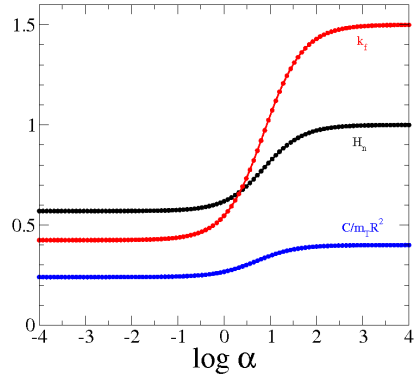}
\caption{Values of $\mathcal{H}_n$ (black), $k_f$ (red) and $C/m_TR_n^2$ (blue) for different values of the exponent of the polynomial density law.}
\label{fig6}
\end{center}
\end{figure}

Note that, as discussed in Section 3, the value of $\mathcal{H}_n$ is always greater than the limit value 0.4 and less than 1. Particularly 
$\mathcal{H}_n$ tends to 0.570 when $\alpha$ tends to 0 and $\mathcal{H}_n$ tends to 1 when $\alpha$ tends to $\infty$ (homogeneous case). The fluid 
Love number increases from 0.424 (when $\alpha$ tends to 0) to 1.5  (when $\alpha$ tends to $\infty$). These results can be seen in Figure (\ref{fig6}), 
where we also show the values of the flattening factor $\mathcal{H}_n$ at the surface and the dimensionless moment of inertia $C/m_TR_n^2$. This last 
parameter increases from 0.24 (when $\alpha$ tends to 0) to 0.4 (when $\alpha$ tends to $\infty$)\footnote{An elementary calculation allows 
one to find the relationship $\frac{C}{m_TR^2}\approx \frac{2}{3}\frac{\int_0^1 \widehat{\rho}z^4  dz}{\int_0^1 \widehat{\rho}z^2dz}=\frac{2}{5}\times\frac{3+\alpha}{5+\alpha}$.}.

\subsection{Polytropics pressure-density laws}
\label{sec:gases ideales}

We may consider a self-gravitating body in hydrostatic equilibrium with a more general polytropic pressure-density law:
\begin{equation}
P=K\rho^{1+\frac{1}{n}},
\end{equation}
where $P$ is the pressure, $n$ is the polytropic index and $K$ is constant. The differential equation for the density is then given by the 
\emph{Lane-Emden equation} (Chandrasekhar, 1939)
\begin{equation}
 \displaystyle\frac{1}{\xi^2}\frac{d}{d\xi}\left(x^2\frac{d\theta}{d\xi}\right)+\theta^n=0,
\label{eq:lane-emden}
\end{equation}
where $\widehat{\rho}=\theta^{n}$ and $R=\alpha\xi$ with $\alpha^2=(n+1)K\rho_0^{\frac{1}{n}-1}/4\pi G$. The standard boundary conditions 
are $\theta(0)=1$ and $\theta'(0)=0$. If $0\leq n<5$ the solution $\theta(\xi)$ decreases monotonically and has a zero at a finite value $\xi=\xi_1$. 
This radius corresponds to the surface of the body where $P=\rho=0$.

It is worth mentioning that several real cases exist that correspond to polytropes. For example, when convection is established in the interior of a 
star the resulting configuration is a polytrope; when the gas is degenerate, the corresponding equations of state have the same form as the polytropic 
equation of state, etc. (see Collins, 1989). We also mention recent results by Leconte et al. (2011) showing that the density profile of hot Jupiters is well 
approximated by a polytrope.

\begin{figure} [here]

\centerline{\hbox{
\includegraphics[height=5cm,clip=]{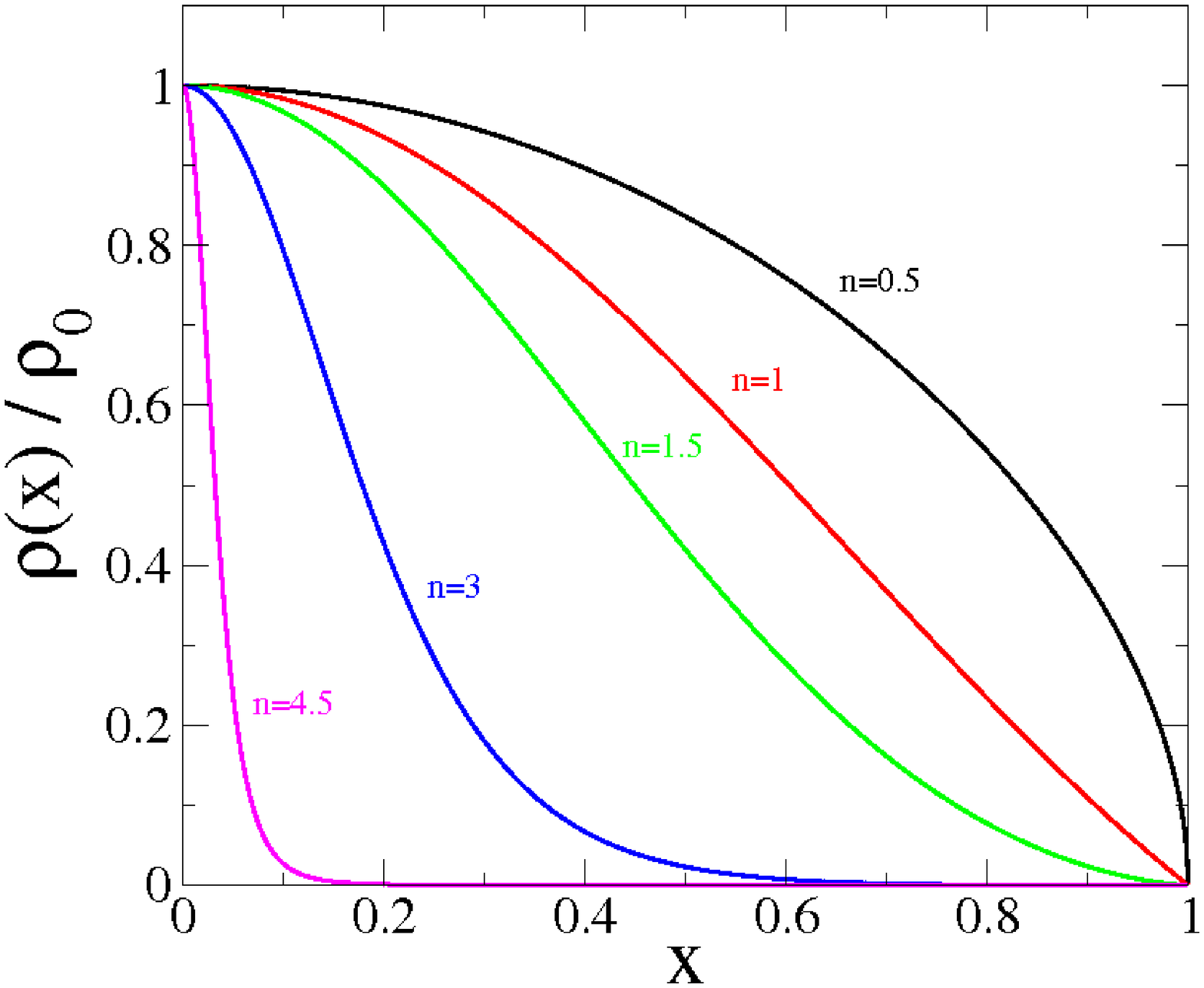}\hspace{3mm}
\includegraphics[height=5cm,clip=]{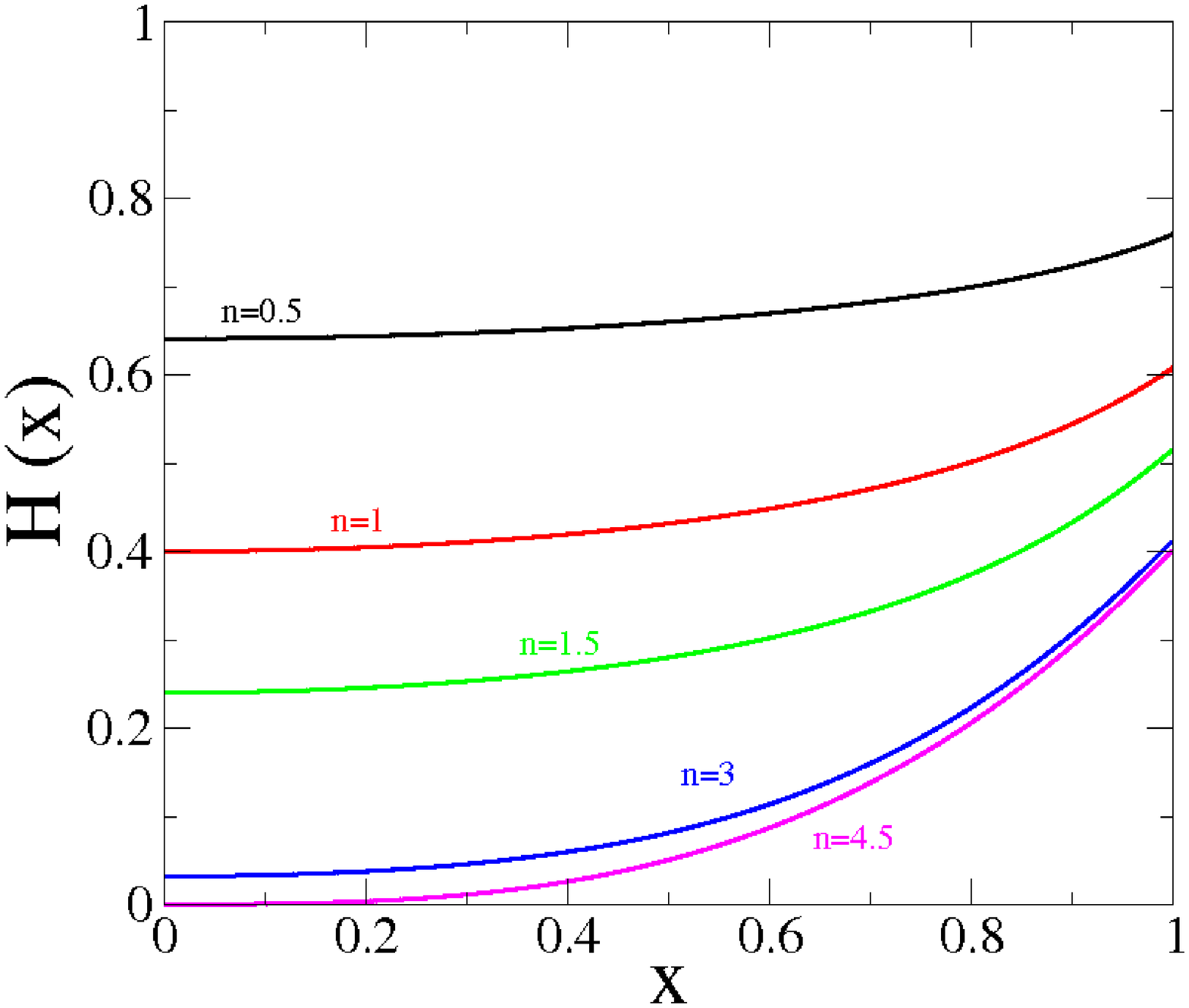}
}}
\begin{center}
\textbf{Fig. \ref{fig7}.a}\ \ \ \ \ \ \ \ \ \ \ \ \ \ \ \ \ \ \ \ \ \ \ \ \ \ \ \ \ \ \ \ \ \ \ \ \ \ \ \ \ \ \ \ \ \ \ \ \  \
\ \ \ \ \ \ \ \ \ \ \ \ \ \ \textbf{Fig. \ref{fig7}.b}

\end{center}
\caption{\textbf{(a)}: Density profiles for different values of the polytropic index. \textbf{(b)}: Flattening profile $\mathcal{H}(x)$ for these 
density laws. $n=0.5$ (black), $n=1$ (red), $n=1.5$ (green), $n=3$ (blue) and $n=4.5$ (magenta).}
\label{fig7}
\end{figure}

Figure (\ref{fig7}.a) shows the density functions for $n=0.5,1.0,1.5,3.0$ and $4.5$ as functions of the normalized mean radius 
$x=R/\alpha\xi_1$ obtained from the integration of the Lane-Emden equation.

The resulting flattening profiles $\mathcal{H}(x)$ are shown in Figure (\ref{fig7}.b). In all cases, the flattening profile $\mathcal{H}(x)$ is 
an increasing monotonic function and for all $x$, the values of $\mathcal{H}(x)$ decrease when the polytropic index $n$ increases.

\begin{figure}[h]
\begin{center}
\includegraphics[height=5cm,clip=]{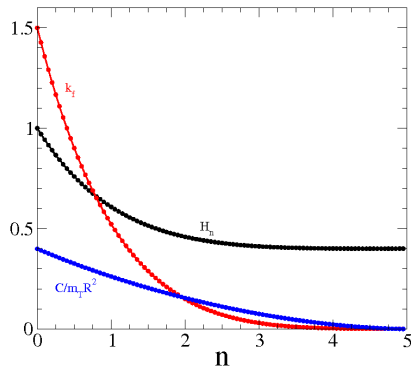}
\caption{Values of $\mathcal{H}_n$ (black), $k_f$ (red) and $C/m_TR_n^2$ (blue) for different polytropic indices $n<5$.}
\label{fig8}
\end{center}
\end{figure}

As mentioned previously, the value of $\mathcal{H}_n$ is always greater than the limit value 0.4. Particularly $\mathcal{H}_n\rightarrow0.4$ 
when $n\rightarrow5$. The fluid Love number decreases from 1.5 for $n=0$ (constant density) to 0 when $n$ tends to the limit $n=5$. These results can 
be seen in Figure (\ref{fig8}), where we also show the values of the flattening factor $\mathcal{H}_n$ and the dimensionless moment of inertia 
$C/m_TR_n^2$ for values of n below the limit $n=5$. The adimensional moment of inertia decreases from 0.4 (when $n=0$) and tends to 0 when $n \rightarrow 5$.

\section{Conclusions}

In this paper, we have extended the classical results on non-homogeneous rotating figures of equilibrium to the case in which the body is also under 
the action of a tidal potential due to the presence of an external body, without the restrictive hypothesis of spin-orbit synchronization. The only 
assumptions in this paper are a body formed by $n$ homogeneous ellipsoidal layers in equilibrium and small enough tidal and rotational deformations 
with symmetry axes perpendicular to each other (remember that, in the second order, the figure ceases to be an ellipsoid). We have calculated 
the $2n$ equilibrium equations for small flattenings and we have found that the two polar flattenings $\epsilon_k$ and $\mu_k$ are linearly related, 
both being proportional to the homogeneous reference values with a factor of proportionality $\mathcal{H}_k$ which is the same in both cases. The 
deformations propagate towards the interior of the body in the same way depending, in the first approximation, only on the density profile; it does 
not depend on the origin of the two considered deformations. Then the problem of finding the $2n$ flattenings, correspond to finding the $n$ 
coefficients $\mathcal{H}_k$ with $n$ equilibrium equations. An important consequence of this approach is that the flattening profile 
$\mathcal{H}_k$ is the same no matter if the rotation of the body is synchronous or non-synchronous and the results for $\mathcal{H}_k$ are the same found by Tricarico 
(2014).

We have also studied the continuous case as the limit for a very large number of layers of infinitesimal thickness, which leads to the Clairaut's 
differential equation for the function $\mathcal{H}(x)$ (i.e. the same equation for both flattenings). This result was expected because the 
coefficients of the Clairaut equation only depend on the internal distribution of matter $\rho(x)$. Therefore, the differential equation that 
generates the functional form of the profile flattening $\mathcal{H}(x)$ does not change when we change the nature of the deformation, provided that 
it is small. For densities decreasing monotonically with the radius, we have found that, at the surface, $\mathcal{H}_n$ takes values larger 
than 0.4 (see Eq. (\ref{eq:h<0.4})) and takes the limit value 1 in the homogeneous case. This means that the surface flattenings of a differentiated 
body are always smaller than the flattening of the corresponding homogeneous ellipsoids, but always larger than 40\% of it.

The results were applied to several examples. In the case of a body composed of two homogeneous layers the following results were obtained:
\begin{itemize}
 \item In a realistic case where the core is denser than the shell, the flattening of the nucleus is smaller than the flattening of the surface. 
This is a result classically known to Tisserand (1891) and discussed in recent papers by Zharkhov and Trubitsyn (1978), Hubbard (2013), 
Tricarico (2014).
 \item In the presence of a core, the surface is always less flattened than the homogeneous reference flattening, but larger than 40\% of the 
latter value.
 \item The fluid Love number $k_f<1.5$, define a continuous curve of possible values for the size of the nucleus $\xi$ and the relative density 
of the shell $\lambda$ and predicts their maximum value.
\end{itemize}

Finally, we studied bodies with different continuous density laws, first for some simple polynomial functions and then for polytropic 
profiles. The following results were obtained:
\begin{itemize}
 \item In all cases, the function $\mathcal{H}(x)$ is an increasing monotonic function.
 \item For all $x$, the values of $\mathcal{H}(x)$ increase from 0.530 to 1 when the power $\alpha$ increases from 0 to $\infty$, in 
contrast with the polytropic densities, in which the values of $\mathcal{H}(x)$ decrease from 1 to 0.4 when the polytropic index $n$ increases from 0 
to the limit case $n=5$.
  \item The fluid Love number $k_f$ varies between 0.326 and 1.5 in the same range of the power $\alpha$ for polynomial densities. For 
the polytropic laws, the fluid Love number $k_f$ varies between from 1.5 to 0 when the polytrocic index $n$ increases.
 \item For polynomial laws, the values of $C/m_TR^2$ increases from 0.24 to 0.4 when the power $\alpha$ increases and for the polytropic 
laws, the values of $C/m_TR^2$ decreases from 0.4 to 0 when the polytropic index $n$ increases.
\end{itemize}

\begin{acknowledgements}
The authors wish to thank one anonymous referee for comments and suggestions that helped to improve the manuscript. This investigation was 
supported by CNPq, grants 141684/2013-5 and 306146/2010-0, and by St. Petersburg University, grant 6.37.341.2015.
\end{acknowledgements}

\vfill\eject

\section*{Online Supplement}
\appendix 
\renewcommand{\theequation}{A.\arabic{equation}}
\setcounter{equation}{0}  
\section{Gravitational potentials}\label{sec:appA}

\subsection{\textit{Potential of a homogeneous ellipsoid at an internal point}}
\label{apen.a}
Let us consider the contribution by the $k$-th ellipsoid at one point $\vec{x}_\ell=\xi_\ell\hat{x}+\eta_\ell\hat{y}+\zeta_\ell\hat{z}$ on the 
surface of the $\ell$-th ellipsoid, assumed interior to it. The potential of these ellipsoid may be written as
\begin{equation}
 V_k(\xi_\ell,\eta_\ell,\zeta_\ell) = V_{0k} + A_\xi \xi_\ell^2 + A_\eta \eta_\ell^2 + A_\zeta \zeta_\ell^2,
\label{eq:Vint}
\end{equation}
where the coefficients $V_{0k}$, $A_\xi$, $A_\eta$ and $A_\zeta$ are
\begin{eqnarray}
 V_{0k}  &=& - \pi G \sigma_k c^2_k \int_0^\infty \displaystyle\frac{dt}{\sqrt{(1+\alpha_k t)(1+\beta_k t)(1+t)}} \nonumber \\
 A_\xi   &=& \ \pi G \sigma_k \alpha_k \int_0^\infty \displaystyle\frac{dt}{(1+\alpha_k t)\sqrt{(1+\alpha_k t)(1+\beta_k t)(1+t)}} \nonumber \\
 A_\eta  &=& \ \pi G \sigma_k \beta_k  \int_0^\infty \displaystyle\frac{dt}{(1+\beta_k t) \sqrt{(1+\alpha_k t)(1+\beta_k t)(1+t)}} \nonumber \\
 A_\zeta &=& \ \pi G \sigma_k \gamma_k \int_0^\infty \displaystyle\frac{dt}{(1+\gamma_k t)\sqrt{(1+\alpha_k t)(1+\beta_k t)(1+t)}},
\end{eqnarray}
and $G$ is the gravitational constant (see Tisserand, 1891, Chap. 8 and 13; Jardetzky, 1958, Sec. 2.2). Then the derivatives of the potential 
are
\begin{eqnarray}
 \displaystyle\frac{1}{\xi_\ell}  \displaystyle\frac{\partial V_k}{\partial \xi_\ell}   &=&  2\pi G \sigma_k \alpha_k \int_0^\infty \displaystyle\frac{dt}{(1+\alpha_k t)^{3/2}(1+\beta_k t)^{1/2}(1+t)^{1/2}} \nonumber \\
 \displaystyle\frac{1}{\eta_\ell} \displaystyle\frac{\partial V_k}{\partial \eta_\ell}  &=&  2\pi G \sigma_k \beta_k  \int_0^\infty \displaystyle\frac{dt}{(1+\alpha_k t)^{1/2}(1+\beta_k t)^{3/2}(1+t)^{1/2}}  \nonumber \\
 \displaystyle\frac{1}{\zeta_\ell}\displaystyle\frac{\partial V_k}{\partial \zeta_\ell} &=&  2\pi G \sigma_k          \int_0^\infty \displaystyle\frac{dt}{(1+\alpha_k t)^{1/2}(1+\beta_k t)^{1/2}(1+t)^{3/2}},
\end{eqnarray}
and its contribution to the $\ell$-th equation of equilibrium are
\begin{eqnarray}
 \chi_\ell^{(1)}(V_k) &=& \displaystyle\frac{1}{\xi_\ell}\displaystyle\frac{\partial V_k}{\partial \xi_\ell}- \displaystyle\frac{\alpha_\ell}{\zeta_\ell}\displaystyle\frac{\partial V_k}{\partial \zeta_\ell} = 2A_\xi-2\alpha_\ell A_\zeta\nonumber\\
                      &=& 2\pi G\sigma_k \int_0^\infty \displaystyle\frac{\alpha_k(1-\alpha_\ell)t+(\alpha_k-\alpha_\ell)}{(1+\alpha_k t)^{3/2}(1+\beta_k t)^{1/2}(1+t)^{3/2}}dt.
\end{eqnarray}

Neglecting terms of order 2 in the flattenings we can write
\begin{equation}
 \chi_\ell^{(1)}(V_k) = 2\pi G\sigma_k \Big[ (1-\alpha_\ell) S_1(\alpha_k,\beta_k)+(\alpha_k-\alpha_\ell)S_2(\alpha_k,\beta_k)\Big],
\end{equation}
where $S_i$ are functions of $\alpha_k$ and $\beta_k$. If we expand around $\alpha_k,\beta_k \sim 1$
\begin{eqnarray}
 S_1(\alpha_k,\beta_k) &=& \int_0^\infty\displaystyle\frac{t}{(1+\alpha_k t)^{3/2}(1+\beta_k t)^{1/2}(1+t)^{3/2}}dt \nonumber\\
                      &\backsimeq& \frac{4}{15} + \frac{8}{35}(1-\alpha_k) + \frac{8}{105}(1-\beta_k) + \dots\nonumber\\
 S_2(\alpha_k,\beta_k) &=& \int_0^\infty\displaystyle\frac{1}{(1+\alpha_k t)^{3/2}(1+\beta_k t)^{1/2}(1+t)^{3/2}}dt \nonumber\\
                      &\backsimeq& \frac{2}{5} + \frac{6}{35}(1-\alpha_k) + \frac{2}{35}(1-\beta_k) + \dots.
\end{eqnarray}

Finally,  we obtain
\begin{equation}
 \chi_\ell^{(1)}(V_k) = \frac{Gm_k}{R_k^3} \left[2\epsilon_\ell-\displaystyle\frac{6\epsilon_k}{5}\right].
\label{eq:int1}
\end{equation}

If $k=\ell$ we have
\begin{equation}
 \chi_\ell^{(1)}(V_\ell) = \frac{Gm_\ell}{R_\ell^3} \frac{4\epsilon_\ell}{5},
\label{eq:l1}
\end{equation}
and similar equation for $\chi_\ell^{(2)}$.

\subsection{\textit{Potential of a homogeneous ellipsoid at an external point}}

The potential generated by the $k$-th homogeneous ellipsoid  at an external point $\vec{x}_\ell=\xi_\ell\hat{x}+\eta_\ell\hat{y}+\zeta_\ell\hat{z}$  
on the surface of the $\ell$-th layer may be presented by Laplace series. Neglecting harmonics of degree higher than 2 we have
\begin{equation}
 V_k(\xi_\ell,\eta_\ell,\zeta_\ell) = -\displaystyle\frac{Gm_k}{r^*} - \displaystyle\frac{GI_k}{r^{*3}} + \displaystyle\frac{3G}{2r^{*5}}\left[I_{\xi\xi}\xi_\ell^2+I_{\eta\eta}\eta_\ell^2+I_{\zeta\zeta}\zeta_\ell^2+2I_{\xi\eta}\xi_\ell\eta_\ell+2I_{\eta\zeta}\eta_\ell\zeta_\ell+2I_{\zeta\xi}\zeta_\ell\xi_\ell\right],
\end{equation}
where $r^*=|\vec{x}_\ell|$, $I_k$ is the moment of inertia of the $k$-th ellipsoid, relative to the center of mass and $I_{ij}$ are the 
components of its inertia tensor (see Beutler, 2005; Murray and Dermott, 1999). If the reference axes are oriented following the principal axes 
of inertia, then $I_{ij}=0$ if $i\neq j$. We may define $A_k=I_{\xi\xi}$, $B_k=I_{\eta\eta}$ and $C_k=I_{\zeta\zeta}$. Because $2I_k=A_k+B_k+C_k$
\begin{equation}
 V_k(\xi_\ell,\eta_\ell,\zeta_\ell) = -\displaystyle\frac{Gm_k}{r^*} - \frac{G(C_k-A_k)}{2r^{*5}}\left(3\xi_\ell^2-r^{*2}\right) - \frac{G(C_k-B_k)}{2r^{*5}}\left(3\eta_\ell^2-r^{*2}\right).
\end{equation}

The principal moments of inertia of a homogeneous ellipsoid are
\begin{eqnarray}
 A_k &=& \frac{1}{5}m_k(b_k^2+c_k^2)=\frac{1}{5}m_kc_k^2\left(\frac{1}{\beta_k}+1\right) \nonumber\\
 B_k &=& \frac{1}{5}m_k(a_k^2+c_k^2)=\frac{1}{5}m_kc_k^2\left(\frac{1}{\alpha_k}+1\right) \nonumber\\
 C_k &=& \frac{1}{5}m_k(a_k^2+b_k^2)=\frac{1}{5}m_kc_k^2\left(\frac{1}{\alpha_k}+\frac{1}{\beta_k}\right) \rightarrow \frac{1}{5}m_kc_k^2=\frac{\alpha_k\beta_k}{\alpha_k+\beta_k}C_k,
\end{eqnarray}
so its subtraction can be approximated
\begin{eqnarray}
 C_k-A_k &=& \frac{(1-\alpha_k)\beta_k}{\alpha_k+\beta_k}C_k\approx\epsilon_k C_k \approx \frac{2}{5}m_kR_k^2\epsilon_k\nonumber\\
 C_k-B_k &=& \frac{(1-\beta_k)\alpha_k}{\alpha_k+\beta_k}C_k\approx\mu_k C_k \approx \frac{2}{5}m_kR_k^2\mu_k.
\end{eqnarray}
therefore, the potential can be written as
\begin{equation}
 V_k(\xi_\ell,\eta_\ell,\zeta_\ell) = -\displaystyle\frac{Gm_k}{r^*} - \frac{Gm_kR_k^2\epsilon_k}{5r^{*5}}\left(3\xi_\ell^2-r^{*2}\right) - \frac{Gm_kR_k^2\mu_k}{5r^{*5}}\left(3\eta_\ell^2-r^{*2}\right),
\label{eq:Vext}
\end{equation}
and its derivatives are
\begin{eqnarray}
 \displaystyle\frac{1}{\xi_\ell}   \frac{\partial V_k}{\partial \xi_\ell}    &=& \frac{Gm_k}{r^{*3}} + \frac{2Gm_kR_k^2}{5r^{*5}}(-2\epsilon_k+\mu_k) + \frac{Gm_kR_k^2\epsilon_k}{r^{*7}} \left(3\xi_\ell^2-r^{*2}\right) + \frac{Gm_kR_k^2\mu_k}{r^{*7}} \left(3\eta_\ell^2-r^{*2}\right) \nonumber \\
 \displaystyle\frac{1}{\eta_\ell}  \frac{\partial V_k}{\partial \eta_\ell}   &=& \frac{Gm_k}{r^{*3}} + \frac{2Gm_kR_k^2}{5r^{*5}}(\epsilon_k-2\mu_k)  + \frac{Gm_kR_k^2\epsilon_k}{r^{*7}} \left(3\xi_\ell^2-r^{*2}\right) + \frac{Gm_kR_k^2\mu_k}{r^{*7}} \left(3\eta_\ell^2-r^{*2}\right) \nonumber \\
 \displaystyle\frac{1}{\zeta_\ell} \frac{\partial V_k}{\partial \zeta_\ell}  &=& \frac{Gm_k}{r^{*3}} + \frac{2Gm_kR_k^2}{5r^{*5}}(\epsilon_k+\mu_k)   + \frac{Gm_kR_k^2\epsilon_k}{r^{*7}} \left(3\xi_\ell^2-r^{*2}\right) + \frac{Gm_kR_k^2\mu_k}{r^{*7}} \left(3\eta_\ell^2-r^{*2}\right).\nonumber\\
\end{eqnarray}

Then we obtain
\begin{eqnarray}
 \chi_\ell^{(1)}(V_k) &=& \displaystyle\frac{1}{\xi_\ell}\frac{\partial V_k}{\partial \xi_\ell}- \frac{\alpha_\ell}{\zeta_\ell}\frac{\partial V_k}{\partial \zeta_\ell}=\frac{1}{\xi_\ell}\frac{\partial V_k}{\partial \xi_\ell}- \frac{1}{\zeta_\ell}\frac{\partial V_k}{\partial \zeta_\ell} + \frac{2\epsilon_\ell}{\zeta_\ell}\frac{\partial V_k}{\partial \zeta_\ell} \nonumber \\
                &\approx& -\displaystyle\frac{6Gm_kR_k^2}{5r^{*5}}\epsilon_k + \frac{2Gm_k}{r^{*3}}\epsilon_\ell,
\end{eqnarray}
where we have neglected the higher-order products $\epsilon_k\epsilon_\ell$ and $\epsilon_\ell\mu_k$. Finally, by making the approximation $r^*\backsimeq R_\ell$, we obtain
\begin{equation}
 \chi_\ell^{(1)}(V_k) = \frac{Gm_k}{R_\ell^3} \left[ 2\epsilon_\ell - \frac{6\epsilon_k}{5}\left(\frac{R_k}{R_\ell}\right)^2\right],
\end{equation}
and, in the same way as before
\begin{equation}
 \chi_\ell^{(2)}(V_k) = \frac{Gm_k}{R_\ell^3} \left[ 2\mu_\ell - \frac{6\mu_k}{5}\left(\frac{R_k}{R_\ell}\right)^2\right]. 
\end{equation}

\subsection{\textit{Tidal potential}}
If $\vec{r}=r\hat{x}$ is the position of the mass M, the tidal potential at a point $\vec{x}_\ell=\xi_\ell\hat{x}+\eta_\ell\hat{y}+\zeta_\ell\hat{z}$  
on the surface of the $\ell$-th layer is (Lambeck 1980)
\begin{equation}
 V_{tid} = -\frac{GM r^{*2}}{r^3} P_2(\vec{\hat{r}}\cdot\vec{\hat{x}}_\ell)
\end{equation}
where $r^*=|\vec{x}_\ell|\backsimeq R_\ell$ and $P_2$ is the Legendre polynomial of degree two. The differential acceleration of this point is
\begin{equation}
 -\nabla V_{tid} = -\displaystyle\frac{GM}{r^3}\left[\vec{x}_\ell-\displaystyle\frac{3(\vec{x}_\ell \cdot \vec{r})}{r^2}\vec{r}\right]=-\displaystyle\frac{GM}{r^3}\left[\begin{array}{c}
-2\xi_\ell\\
\eta_\ell\\
\zeta_\ell \end{array}\right],
\end{equation}
therefore their derivatives are
\begin{eqnarray}
 \displaystyle\frac{1}{\xi_\ell}  \displaystyle\frac{\partial V_{tid}}{\partial \xi_\ell}   &=& -\displaystyle\frac{2GM}{r^3}\nonumber \\
 \displaystyle\frac{1}{\eta_\ell} \displaystyle\frac{\partial V_{tid}}{\partial \eta_\ell}  &=&  \displaystyle\frac{GM}{r^3}\nonumber\\
 \displaystyle\frac{1}{\zeta_\ell}\displaystyle\frac{\partial V_{tid}}{\partial \zeta_\ell} &=&  \displaystyle\frac{GM}{r^3}.
\end{eqnarray}

Finally, the contribution of the tide in the equilibrium equations of the $\ell$-th layer is
\begin{eqnarray}
 \chi_\ell^{(1)}(V_{tid}) &=& \displaystyle\frac{1}{\xi_\ell}\displaystyle\frac{\partial V_{tid}}{\partial \xi_\ell}- \displaystyle\frac{\alpha_\ell}{\zeta_\ell}\displaystyle\frac{\partial V_{tid}}{\partial \zeta_\ell} \nonumber \\
&=& -\displaystyle\frac{3GM}{r^3}+\displaystyle\frac{2GM}{r^3}\epsilon_\ell,
\end{eqnarray}
and, similarly,
\begin{eqnarray}
 \chi_\ell^{(2)}(V_{tid}) &=& \displaystyle\frac{1}{\eta_\ell}\displaystyle\frac{\partial V_{tid}}{\partial \eta_\ell}- \displaystyle\frac{\beta_\ell}{\zeta_\ell}\displaystyle\frac{\partial V_{tid}}{\partial \zeta_\ell} \nonumber \\
&=& \displaystyle\frac{GM}{r^3}\mu_\ell.
\end{eqnarray}

However, we discard terms containing $\epsilon_\ell$ and $\mu_\ell$, because when we calculate the flattenings of each layer, they appear multiplied by a 
factor of the same order as $\epsilon$ or $\mu$, therefore we obtain
\begin{equation}
 \chi_\ell^{(1)}(V_{tid}) = -\displaystyle\frac{3GM}{r^3},
\end{equation}
and
\begin{equation}
 \chi_\ell^{(2)}(V_{tid}) = 0.
\end{equation}

\renewcommand{\theequation}{B.\arabic{equation}}
\setcounter{equation}{0}  
\section{MacLaurin, Jeans and Roche ellipsoids}\label{sec:appB}

In this appendix, we show that, in the homogeneous case, the flattenings defined by eqs. (\ref{def:em}) correspond to the MacLaurin and Jeans ellipsoids, 
respectively. In the homogeneous case, the equation (\ref{eq:equi}) takes the simple form
\begin{equation}
 \chi_1^{(i)}(V) = \chi_1^{(i)}(V_{tid}) + \chi_1^{(i)}(V_1)= \Omega^2,
\end{equation}
In this case, the contribution of the attraction of this ellipsoid to the equilibrium equation is
\begin{eqnarray}
 \chi_1^{(1)}(V_1) &=& \frac{Gm}{R^3} \frac{4\epsilon_1}{5},\nonumber\\
 \chi_1^{(2)}(V_1) &=& \frac{Gm}{R^3} \frac{4\mu_1}{5},
\end{eqnarray}
and the tidal contribution is
\begin{eqnarray}
 \chi_\ell^{(1)}(V_{tid}) &=& -\displaystyle\frac{3GM}{r^3},\nonumber\\
 \chi_\ell^{(2)}(V_{tid}) &=& 0.
\end{eqnarray}

The solutions can readily be found. They are
\begin{eqnarray}
 \epsilon_1 &=& \frac{15}{4}\frac{M}{m}\left(\frac{R}{r}\right)^3+\frac{5R^3\Omega^2}{4mG},\nonumber\\
 \mu_1      &=& \frac{5R^3\Omega^2}{4mG}.
\label{def:sh}
\end{eqnarray}

\begin{figure}[here]
\begin{center}
\includegraphics[scale=0.3]{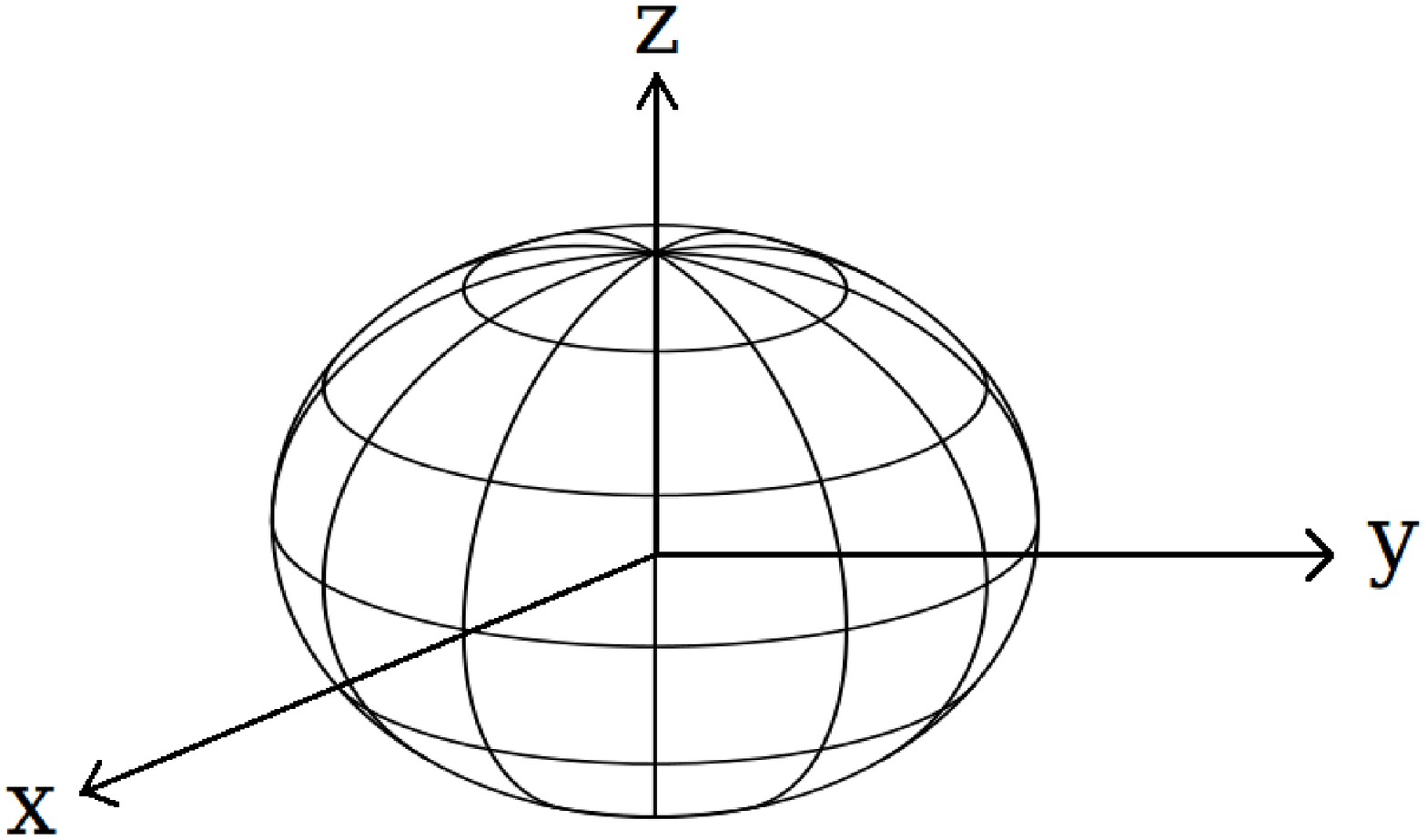}
\includegraphics[scale=0.3]{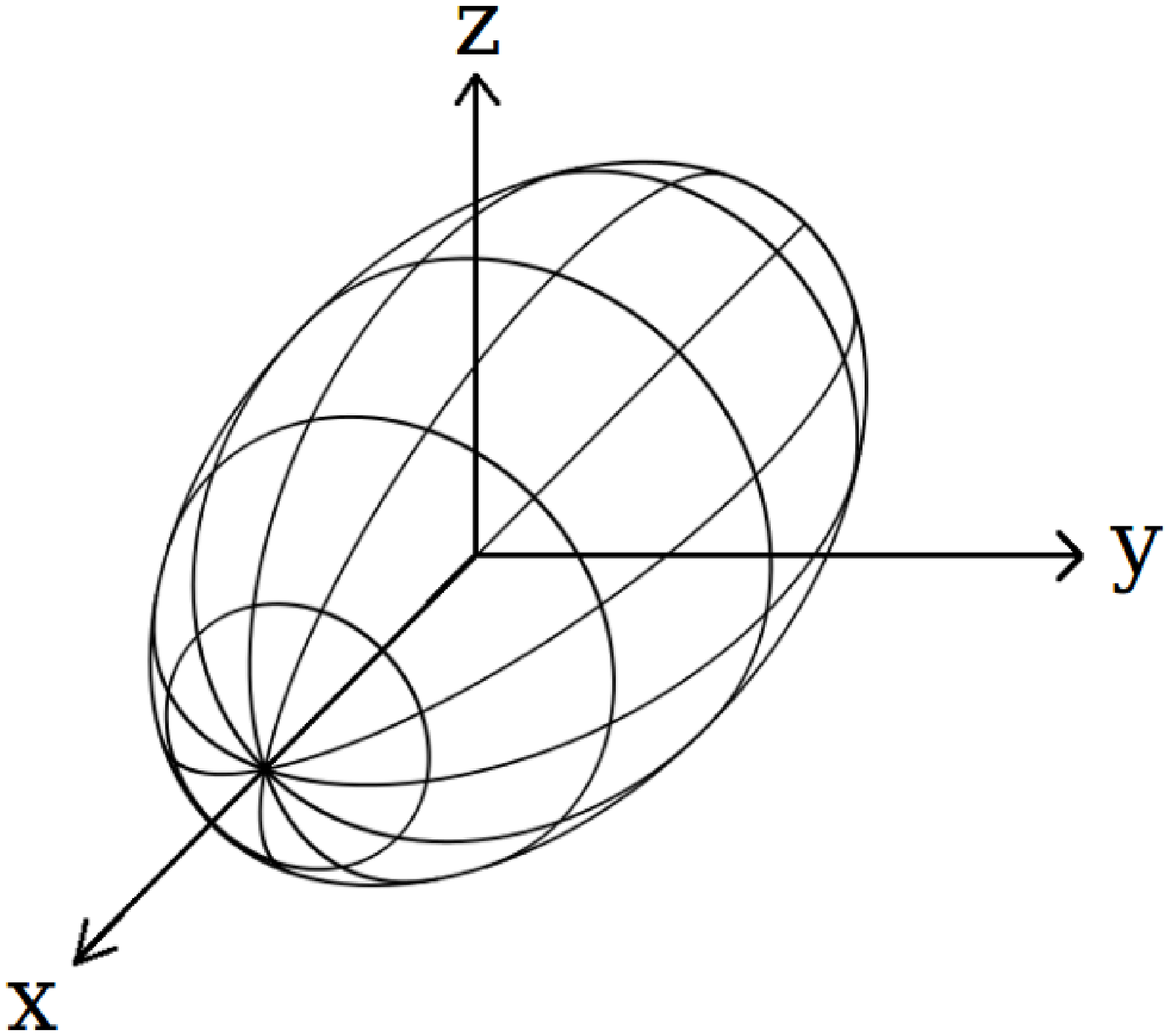}
\\ \textbf{Fig. \ref{fig0}.a} \ \ \ \ \ \ \ \ \ \ \ \ \ \ \ \ \ \ \ \ 
\ \ \ \ \ \ \ \ \ \ \ \ \ \ \ \ \ \ \ \   \textbf{Fig. \ref{fig0}.b}
\caption{\textbf{a}) MacLaurin oblate spheroid when the axis of rotation is on the $z$ axis. In this case, $a=b>c$. \textbf{b}) Jeans prolate spheroid 
when the mass disturbing the particle lies on the axis $x$. In this case, $a>b=c$.}
\label{fig0}
\end{center}
\end{figure}

When the effect of the rotation is the only considered, the homogeneous solution was discovered by Newton and later generalized by MacLaurin 
(Chandrasekhar, 1969). The solution is an oblate spheroid with semiaxes $a=b>c$, with polar flattening $\epsilon_1=\mu_1=\epsilon_M$
\begin{equation}
\epsilon_M = \frac{5R^3\Omega^2}{4mG}.
\end{equation} 
On the other hand, when the effect of the tide is the only considered, the homogeneous solution to the first order is a Roche ellipsoid with two 
equal axes (Tisserand, 1981). This equlibrium ellipsoid was later called Jeans ellipsoid (Chandrasekhar, 1969). It is a prolate 
spheroid with semiaxes $a>b=c$, with $\mu_1=0$ and $\epsilon_1=\epsilon_J$
\begin{equation}
 \epsilon_J = \frac{15}{4}\frac{M}{m}\left(\frac{R}{r}\right)^3.
\end{equation}
Figure \ref{fig0} illustrates the two cases.

When we consider both effects, the solutions (\ref{def:sh}) correspond to a triaxial ellipsoid with semiaxes $a>b>c$ and the polar flattenings are 
$\epsilon_1=\epsilon_J+\epsilon_M$ and $\mu_1=\epsilon_M$.

\renewcommand{\theequation}{B.\arabic{equation}}
\setcounter{equation}{0}  
\section{Extension to the continuous case}\label{sec:appC}

In this appendix, we show the intermediary calculations necessary to find the Clairaut's equation, following Tisserand (1891). Let us start from equations 
(\ref{eq:achata2}), and let us make the transition to the continuum. If we introduce the notation $\Delta(g_k)=g_k-g_{k-1}$ and the boundary values 
$R_0=0$ and $\rho_{n+1}=0$, we may rewrite the terms on the right hand side of eqn. (\ref{eq:achata2}) as
\begin{eqnarray}
 \displaystyle\sum_{k=1}^{\ell-1} \alpha_{\ell k}\mathcal{H}_k  &=& \sum_{k=1}^{\ell-1} \frac{3m_k}{2m_T}\left(\frac{R_k}{R_\ell}\right)^2 \mathcal{H}_k =\displaystyle \frac{2\pi}{m_TR_\ell^2} \sum_{k=1}^{\ell-1} (\rho_k-\rho_{k+1})R_k^5 \mathcal{H}_k\nonumber\\
                                                                &=&  \displaystyle - \frac{2\pi}{m_T}\rho_\ell R_\ell^3\mathcal{H}_\ell +\frac{2\pi}{m_T} \frac{1}{R_\ell^2}\sum_{k=1}^\ell \rho_k \Delta(R_k^5 \mathcal{H}_k)  \nonumber\\
 \displaystyle\sum_{k=\ell+1}^{n} \beta_{\ell k}\mathcal{H}_k   &=& \sum_{k=\ell+1}^{n} \frac{3m_k}{2m_T}\left(\frac{R_\ell}{R_k}\right)^3 \mathcal{H}_k =\displaystyle \frac{2\pi}{m_T}R_\ell^3 \sum_{k=\ell+1}^{n} (\rho_k-\rho_{k+1}) \mathcal{H}_k\nonumber\\
                                                                &=&  \displaystyle \frac{2\pi}{m_T}(\rho_{\ell+1} R_\ell^3\mathcal{H}_\ell-\rho_{\ell} R_\ell^3\Delta(\mathcal{H}_\ell)) +\frac{2\pi}{m_T} \frac{1}{R_\ell^2}\sum_{k=\ell}^{n} \rho_k \Delta( \mathcal{H}_k),
\end{eqnarray}
or 
\begin{eqnarray}
 \displaystyle\sum_{k=1}^{\ell-1} \alpha_{\ell k}\mathcal{H}_k + \sum_{k=\ell+1}^{n} \beta_{\ell k}\mathcal{H}_k &=& \frac{2\pi}{m_T} \left[\frac{1}{R_\ell^2}\sum_{k=1}^\ell \rho_k \Delta(R_k^5 \mathcal{H}_k)+\frac{2\pi}{m_T} R_\ell^3\sum_{k=\ell}^{n} \rho_k \Delta( \mathcal{H}_k)\right]\nonumber\\
       && + \displaystyle \frac{2\pi}{m_T}R_\ell^3(\Delta(\rho_{\ell+1}) \mathcal{H}_\ell-\rho_{\ell} \Delta(\mathcal{H}_\ell)).
\end{eqnarray}

We also have
\begin{equation}
\gamma_\ell = \displaystyle\frac{5}{2} + \frac{2\pi}{m_T} R_\ell^3 \Delta\rho_\ell - \frac{10\pi}{3m_T} \left[\sum_{k=\ell}^n \rho_k\Delta(R_k^3) - \rho_\ell\Delta(R_\ell^3)\right].
\end{equation}

In order to extend to the continuous case, we assume that the number of layers tends to infinity so that the increments $\Delta g$ are 
infinitesimal quantities. When $\Delta g\rightarrow 0$, all terms in which the infinitesimals appear alone tends to zero. Only the terms in which 
they are integrated over an infinite number of layers remain. Hence,
\begin{eqnarray}
\displaystyle\sum_{k=1}^{\ell-1} \alpha_{\ell k}\mathcal{H}_k+\displaystyle\sum_{k=\ell+1}^n \beta_{\ell k}\mathcal{H}_k &\rightarrow& \displaystyle \frac{2\pi}{m_T}\left[ \frac{1}{R^2}\int_{s=0}^{s=R} \rho(s) d\left(s^5 \mathcal{H}(s)\right) + R^3\int_{s=R}^{s=R_n} \rho(s) d\mathcal{H}(s)\right] \nonumber\\
\gamma_\ell                                                                             &\rightarrow& \frac{5m(R)}{2m_T}=\gamma(R),
\end{eqnarray}
where $R_n$ is the mean radius of the ellipsoid and $m(R)$ is the total mass enclosed by an ellipsoid of mean radius $R$. We thus have the integral 
equation
\begin{equation}
 \gamma(R)\mathcal{H}(R) = \displaystyle\frac{R^3}{R_n^3} + \frac{2\pi}{m_T}\left[ \frac{1}{R^2}\int_{s=0}^{s=R} \rho(s)d\left(s^5 \mathcal{H}(s)\right) + R^3\int_{s=R}^{s=R_n} \rho(s) d\mathcal{H}(s)\right].
\label{eq.ach.int}
\end{equation}

It is worth noting that the only restriction for the density profile is that it may be a piecewise continuous function. In order to normalize the 
problem, we first define a new variable
\begin{equation}
 x=\frac{R}{R_n},
\end{equation}
so that $x(0)=0$ in the center and $x(R_n)=1$ on the surface. 

In the next step, we normalize the densities law
\begin{equation}
 \rho(R) = \rho_0\widehat{\rho}(x),
\end{equation}
where $\rho_0$ is the density in the center and $\widehat{\rho}(0)=1$.

At last, we define the function
\begin{equation}
f(x)=3\int_0^{x}\widehat{\rho}(z)z^2dz,
\end{equation}
where $f(0)=0$ and $f(1)=f_n$. The mass enclosed in the $R$-sphere is then
\begin{equation}
 m(R)= 4\pi\int_0^{R}\rho(r)r^2dr = \frac{4\pi}{3} R_n^3\rho_0 f(x),
\end{equation}
and the total mass is
\begin{equation}
 m_T = \frac{4\pi}{3} R_n^3\rho_0f_n.
\end{equation}

Then, the integral equation (\ref{eq.ach.int}) becomes
\begin{equation}
 \frac{5x^2}{3}f(x) \mathcal{H}(x) = \frac{2f_n}{3} x^5 + \int_{z=0}^{z=x} \widehat{\rho}(z) d(z^5\mathcal{H}(z)) + x^5 \int_{z=x}^{z=1} \widehat{\rho}(z) d\mathcal{H}(z),
\end{equation}
or, deriving with respect to $x$ and dividing by $5x^4$,
\begin{equation}
 \frac{2f(x)}{3x^3}\mathcal{H}(x)+\frac{f(x)}{3x^2}\mathcal{H}'(x) = \frac{2f_n}{3} +  \int_{z=x}^{z=1} \widehat{\rho}(z)d\mathcal{H}(z).
\end{equation}

Deriving once more we obtain the Clairaut's equation for the flattening profile $\mathcal{H}(x)$
\begin{equation}
 \mathcal{H}''(x)+ \frac{6\widehat{\rho}(x) x^2}{f(x)} \mathcal{H}'(x) + \left(\frac{6\widehat{\rho}(x) x}{f(x)}-\frac{6}{x^2}\right) \mathcal{H}(x) = 0.
\end{equation}


\begin{thebibliography}{99}

\bibitem{bizyaev.2014}
Bizyaev, I.A., Borisov, A.V. and Mamaev, I.S.: 2014, \textquotedblleft Figures of equilibrium of an inhomogeneous self-gravitating fluid.
\textquotedblright Nelineinaya Dinamika. \textbf{10}, 73–100. (In Russian)

\bibitem{borisov.2009}
Borisov, A.V., Mamaev, I.S. and Kilin, A.A.: 2009, \textquotedblleft The Hamiltonian dynamics of self-gravitating liquid and gas ellipsoids.
\textquotedblright Regul. Chaotic Dyn. \textbf{14}, 179-217.

\bibitem{bullen.1975}
Bullen, K.E.: 1975, \textquotedblleft The Earth's density\textquotedblright, (Chapman and Hall, London).

\bibitem{chandra.1969}
Chandrasekhar, S.: 1969, \textquotedblleft Ellipsoidal Figures of Equilibrium\textquotedblright, (Yale Univ. Press, New Haven).

\bibitem{clairaut.1743}
Clairaut, A.C.: 1743, \textquotedblleft Th\'eorie de la Figure de la Terre, Tir\'ee des Principes de l'Hydrostratique\textquotedblright, 
(Paris Courcier, Paris).

\bibitem{collins.1989}
Collins, G.W.: 1989, \textquotedblleft The fundamentals of stellar astrophysics \textquotedblright, (W.H. Freeman and Co., New York)
 
\bibitem{correia.2013}
Correia, A. and Rodr\'iguez, A.: 2013, \textquotedblleft On the equilibrium figure of close-in planets and satellites.\textquotedblright Astrophys. J. \textbf{767}, 128-132.
 
\bibitem{darwin.1880}
Darwin, G.H.: 1880, \textquotedblleft On the secular change in the elements of the orbit of a satellite revolving about a tidally distorted 
planet.\textquotedblright Philos. Trans. \textbf{171}, 713-891. (repr. Scientific Papers, Cambridge, Vol. II, 1908)

\bibitem{esteban.2001}
Esteban, E.P. and Vazquez, S.: 2001, \textquotedblleft Rotating stratified heterogeneous oblate spheroid in newtonian physics.\textquotedblright 
Celest. Mech. Dyn. Astron. \textbf{81}, 299-312.

\bibitem{ferraz.2008}
Ferraz-Mello, S., Rodr\'iguez, A. and Hussmann, H.: 2008, \textquotedblleft Tidal frition in close-in satellites and exoplanets. The Darwin 
theory re-visited.\textquotedblright Celest. Mech. Dyn. Astron. \textbf{101}, 171-201. Errata: \textbf{104}, 319-320.

\bibitem{ferraz.2013}
Ferraz-Mello, S.: 2013, \textquotedblleft Tidal synchronization of close-in satellites and exoplanets. A rheophysical approach.\textquotedblright 
Celest. Mech. Dyn. Astron. \textbf{116}, 109-140.

\bibitem{gavrilov.1976}
Gavrilov, S.V., Zharkov, V.N. and Leontev, V.V.: 1976, \textquotedblleft Influence of tides on the gravitational field of Jupiter.\textquotedblright 
Soviet Astronomy, \textbf{19}, 618-621.

\bibitem{hubbard.2013}
Hubbard, W.B.: 2013, \textquotedblleft Concentric Maclaurin spheroid models of rotating liquid planets.\textquotedblright 
Astrophys. J. \textbf{768}, 43.

\bibitem{jardetzky.1958}
Jardetzky, W.S.: 1958, \textquotedblleft Theories of Figures of Celestial Bodies\textquotedblright, (Interscience Publ. New York; repr. Dover, 
Mineola, NY, 2005).

\bibitem{jeans.1929}
Jeans, J.: 1929, \textquotedblleft Astronomy and Cosmogony\textquotedblright (Cambridge Univ. Press, Cambridge; repr. Dover, New York, 1961)

\bibitem{jeffreys.1953}
Jeffreys, H.S.: 1953, \textquotedblleft The figures of rotating planets.\textquotedblright Mon. Not. R. astr. Soc. \textbf{113}, 97.

\bibitem{kong.2010}
Kong, D., Zhang, K. and Schubert, G.: 2010, \textquotedblleft Shapes of two-layer models of rotating planets.\textquotedblright 
J. Geophys. Res. \textbf{115}, 12003.

\bibitem{leconte.2011}
Leconte, J., Lai, D. and Chabrier, G.: 2011, \textquotedblleft Distorted, non-spherical transiting planets: impact on the transit depth and on the 
radius determination.\textquotedblright Astronomy \& Astrophysics \textbf{528}, A41. Erratum: Astronomy \& Astrophysics, \textbf{536}, C1.

\bibitem{lyapounov.1925}
Lyapounov, A.: \textquotedblleft Sur certaines s\'eries de figures d'equilibre d'un liquide h\'eterog\`ene en rotation.\textit{Acad. Sci. URSS}, Part 
I, 1925 and Part II 1927.

\bibitem{montalvo.1983}
Montalvo, D., Mart\'inez, F.J. and Cisneros, J.: 1983, \textquotedblleft On equilibrium figures of ideal fluids in the form of confocal spheroids 
rotating with common and different angular velocities.\textquotedblright Rev. Mexicana Astron. Astrof. \textbf{5}, 293-300.

\bibitem{munk.1960}
Munk, W.H. and MacDonald, G.J.F.: 1960, \textquotedblleft The Rotation of the Earth: A Geophysical Discussion\textquotedblright, 
(Cambridge Univ. Press, Cambridge, 1960).

\bibitem{poincare.1902}
Poincar\'e, H.: 1902, \textquotedblleft Figures d'equilibre d'una masse fluide \textquotedblright (Le\c cons profess\'ees \`a la Sorbenne en 1900) 
\textit{Paris, Gauthier-Villars}.

\bibitem{tisserand.1891}
Tisserand, F.: 1891, \textquotedblleft Trait\'e de M\'ecanique C\'eleste\textquotedblright, Tome II, (Gauthier-Villars, Paris).

\bibitem{tricarico.2014}
Tricarico, P.: 2014, \textquotedblleft Multi-layer hidrostatic equilibrium of planets and synchronous moons: Theory and application to Ceres and 
Solar System moons.\textquotedblright Astrophys. J. \textbf{782}, 12.

\bibitem{vanhooslt.2008}
Van Hoolst, T., Rambaux, N., Karatekin, \"{O}., Dehant, V. and Rivoldini, A.: 2008, \textquotedblleft The librations, shape, and icy shell of Europa.
\textquotedblright Icarus \textbf{195}, 386-399.

\bibitem{wavre.1932}
Wavre, R.: 1932, \textquotedblleft Figures plan\'etaries et G\'eodesie \textquotedblright, (Gauthier-Villars et cie, Paris).

\bibitem{zharkov.1978}
Zharkov, V.N. and Trubitsyn, V.P.: 1978, \textquotedblleft Figures plan\'etaries et G\'eodesie \textquotedblright, (Astronomy and Astrophysics Series, 
Tucson: Pachart 1978).

\end{thebibliography}

\begin{thebibliography}{99}
 
\bibitem{Beutler.2005}
Beutler, G.: 2005, \textquotedblleft Methods of Celestial Mechanics\textquotedblright, Vol. I, (Springer, Berlin).

\bibitem{chandra.1969a}
Chandrasekhar, S.: 1969, \textquotedblleft Ellipsoidal Figures of Equilibrium\textquotedblright, (Yale Univ. Press, New Haven).

\bibitem{jardetzky.1958a}
Jardetzky, W.S.: 1958, \textquotedblleft Theories of Figures of Celestial Bodies\textquotedblright, (Interscience Publ. New York; repr. Dover, 
Mineola, NY, 2005).

\bibitem{lambeck.1980}
Lambeck, K.: 1980, \textquotedblleft The Earth's Variable Rotation: Geophysical Causes and Consequences\textquotedblright, (Cambridge Univ. Press, Cambridge)

\bibitem{murray.1999}
Murray, C. and Dermott S.: 1999, \textquotedblleft Solar System Dynamics\textquotedblright, (Cambridge University Press).

\bibitem{tisserand.1891a}
Tisserand, F.: 1891, \textquotedblleft Trait\'e de M\'ecanique C\'eleste\textquotedblright, Tome II, (Gauthier-Villars, Paris).

\end{thebibliography}
\end{document}